\documentclass{PoS}
\usepackage{amsmath}
\title{Anomalies from Non-Perturbative Standard Model Effects}

\ShortTitle{Non-perturbative ``anomalies''}

\author{\speaker{Holger Bech Nielsen}\thanks{A footnote may follow.}\\
        Niels Bohr Institutet, Copenhagen \\
        E-mail: \email{hbech@nbi.dk}}

\author{Colin D. Froggatt\\
        Glasgow University\\
        E-mail: \email{colin.froggatt@gla.ac.uk}}

\abstract{We interpret anomalies/deviations from the
Standard Model as being in fact due to non-perturbative effects, because
 the top-yukawa-coupling is after all so large that non-perturbative
effects become important.  Most of the anomalies found have the character of
signaling violation of lepton universality (LUV).

There are four lepton universality violating anomalies known at present,
which we shall fit with one overall scale parameter $K$ to order of magnitude
accuracy. One can look at the picture that we have found - due to the
rather strong top-yukawa-coupling - as a new sector of strongly interacting
particles analogous to QCD inside the Standard Model!

In addition we treat what is presumably also an anomaly: the experimental
value of the direct CP-violation parameter $\epsilon'$ or $\epsilon'/\epsilon$
turns out to be a factor 2 or more larger than the Standard Model prediction.
It can also be fitted well by our model. However if we include in our anomaly model such processes
not involving leptons (but only quarks), it seems superficially that we
obtain very large - and phenomenologically  unacceptable - anomalies in
$K^0\bar{K}^0$ etc. mixing. To rescue our model from this falsification,
we think of the interacting quarks as provided with clouds (on 1/2 TeV scale)
of, in our
picture, the strongly interacting Higgs and top-quark particles, and that
too strongly interacting such clouds cannot penetrate into each other.
Thereby  the interactions and thus the resulting anomalies are damped.
Even with such a crutch for our theory we only barely manage to avoid an
overlarge anomaly in the indirect CP-violation parameter $\epsilon$.
Our modified model also predicts an anomalous contribution to the
$K_L-K_S$ mass difference of the same order of magnitude as the
experimental value, which is consistent with the (statistically insignificant)
deviation from the Standard Model lattice value.}

\FullConference{Corfu Summer Institute 2018 "School and Workshops on Elementary Particle Physics and Gravity"\\
		(CORFU2018)\\
		31 August - 28 September, 2018\\
		Corfu, Greece}

\usepackage[utf8x]{inputenc}

\begin{document}

\maketitle


\section{Introduction}
As we shall list in section \ref{anomalis} there are (about) 5
small, most barely statistically significant, deviations from the
Standard Model. Interestingly enough 4 out of these 5 anomalies
involve violation of lepton universality. That is to say, there is an anomaly
for one lepton flavour rather than for another one, or at least the anomaly
is different for e.g. electron and muon. Precisely such a muon versus
electron difference is the point of ``the $R(K)$ and $R(K^*)$ anomaly in
B-decay'', `the muon magnetic moment anomalous magnetic moment anomaly''
(there is namely no anomaly in the electron anomalous magnetic moment),
and ``the proton radius puzzle'', which consists in the fact that the  radius of the
proton is fitted to be smaller for muonic data than for the usual electronic atomic data.
``The $R(D)$, $R(D^*)$, and $R(J/\Psi)$ in B-decay anomaly'' is a violation
of the universality between the $\tau$-lepton and the lighter leptons,
$\mu$ say. The only one of the treated anomalies, which is not about lepton
universality violation, is ``the $\epsilon'/\epsilon$ anomaly'' concerned with
direct CP-violation.

It is our point to fit all these 5 anomalies as being due to non-pertubative
effects present in the Standard Model - so that strictly speaking the
Standard Model is perfectly o.k. - order of magnitudewise with only one
overall scale parameter, which we call $K$.

We shall see below that the fitting of the very different observed anomalies
works very
well. But we have the problem, that we must ``improve'' on our model, to
avoid it making predictions for outrageous anomalies in neutral meson
mixings, $K^0\bar{K^0}$ and so on.
Ideas for such an improvement are
included below, but it might be easier to take a version of our model in which
we only consider processes with a lepton. If we do so we strictly
speaking sacrifice the ``the $\epsilon'/\epsilon$ anomaly in CP -violation'',
in spite of it in fact fitting very well using the overall scale
parameter $K$-value obtained by fitting to the lepton involving anomalies.

The philosophical point of our fitting of the anomalies with the overall
scale parameter $K$  is that the top-Yukawa coupling in the Standard Model
$g_t =0.93_5$ is close to being of order unity and thus could in principle
be suspected to - much like the strong gauge coupling in QCD $\alpha_S$,
which runs large in the low energy regime - give rise to
non-perturbative effects becoming important. In fact we propose
that the top-Yukawa coupling $g_t$ is so big as
to give rise to a
{\bf ``new strong interaction sector''}, in which then this
top-Yukawa coupling plays an analogous role to the strong gauge coupling
in  QCD. Whether a coupling like $g_t$ should really be considered
sufficiently strong to cause significant non-perturbative effects
- such as strongly bound states, or significant effective interactions
of a more complicated type - does not only depend on the coupling
being sufficiently strong/large but also on how many different
species of particles that couple to each other by means of this coupling.
At the end we shall give some estimates actually ending up with the conclusion that
the $g_t = .935$ is indeed almost sufficient to be expected to
give severe non-perturbative effects.

In the next section \ref{anomalis} we shall list the deviations from
the Standard Model found and review the main experimental results.

Then in the
section \ref{model} we put forward the rules for our model.

 In  section \ref{Fitting} we go through the various anomalies found
and fit our parameter $K$, an overall scale for the non-perturbative
effect.

In section \ref{resumetable} we present in a table the success of our model:
we get almost the same value for our parameter $K$ by fitting whichever
of the anomalies.

In section \ref{strong} we briefly discuss what the ``strong coupling'', which
is the
basis for nonperturbative effects, actually stands for; first it means of course
that we do not in the true sense have any new physics, so that the
Standard Model can still be perfect. However it might also be considered a
support for our long discussed multiple point principle, that there are
several degenerate vacua. Thirdly we estimate whether the value $ g_t =0.935$ is
expected to be sufficiently strong to indeed give non-perturbative
effects; we find that it is close to the border line for being
so strong as to give significant non-perturbative effects.

In section \ref{mixing} we worry about, how to avoid our model being
totally falsified by predicting too large anomalies
in the pseudoscalar meson mixings, such as $K^0 \bar{K}^0$ mixing.

For this purpose we invent a mechanism, that may for very strongly
interacting particles prevent them from penetrating into the clouds of
top and Higgs particles around them and thus damp the predictions
of our model. This penetration story is described in section \ref{Penetration}.

Section \ref{epsilon} is assigned to
a discussion of the prediction of our modified model
of an anomaly in the CP-violating parameter $\epsilon$ measured in the
first CP-violating experiment by Christensen, Cronin et al.
Really this $\epsilon$ parameter seems to fit
well to the Standard Model. Although at first it looks that our predicted anomaly
is dramatic, we explain in this section, that it is not that serious.

In section
\ref{Conclusion} we conclude with a r\'esum\'e and give some
PREdictions for where soon to find further anomalies.

\section{The Anomalies}
\label{anomalis}

We think we should stress that the type of experiments in which one has
found anomalies, in the sense of deviations from the Standard Model
predictions, are of {\em very different character}: The first two anomalies
though are both deviations from lepton universality (LU) in B-meson decays.
But even these two a priori similar deviations differ in fact rather
remarkably by the first one ``$R(K)$ and $R(K^*)$'' corresponding to a
process, that is of extremely low decay rate, because it is a {\em neutral
(weak) current} process with a couple of charged leptons in the final state,
while the next anomaly ``$R(D^*)$ and $R(j/\Psi)$'' concerns a  charged
current  process, thus with a much bigger decay rate.
The muon and electron anomalous magnetic moments are measured with very special
instrumentation and calculated theoretically with extreme accuracy.
The ``Proton radius puzzle'' is really that atomic physics data with
a muon instead of an electron is fitted by a radius for  the proton 4\% smaller
than that obtained from the electronic/usual atomic data.

If we should choose to only consider
the anomalies in
processes with leptons - and in fact lepton universality
violation - we obtain
our reduced model.
We should then ignore the $\epsilon'/\epsilon$ anomaly, but really we
also manage to fit that non-leptonic anomaly very well.

But the truth is that we have to produce a crutch - a story about
the quarks being stopped on the way penetrating into each other
or rather each others surrounding clouds of say top and
Higgs particles - for our model for
treating the non-perturbative effects, in order to avoid
predicting too huge anomalies for the mixing of the neutral
mesons such as $K^0\bar{K}⁰$. Even after such improvement it is
hard to keep the anomaly in especially the indirect CP-breaking
parameter $\epsilon$ sufficiently small. (We assign
section \ref{epsilon} to
this problem).


As the lepton family in our model comes in via a number of
lepton-Higgs-yukawa couplings, we tend to get much bigger anomalies
the bigger the mass of the lepton in question. Therefore we shall with
our model in mind always think of the heaviest lepton involved,
in the comparisons showing lepton universality violation, as being the
lepton flavour carrying the anomaly. When as in most cases the
lack of universality is between muon and electron, we shall therefore
take it that it is the muon process that is anomalous.
It is a confirmation of this point of view that the anomalous magnetic moment
for the muon has an anomaly, while the electron anomalous magnetic moment
fits the Standard Model perfectly. Only in the case number 2 below,
``the $R(D)$, $R(D^*)$, and $R(J/\Psi)$ anomaly in B-decay'', it is not
the muon but the $\tau$-lepton process that shall carry the anomaly
(it is only in this case that the corresponding $\tau$-process has been sufficiently
well measured.)

\subsection{The $R(K)$, and  $R(K^*)$ anomaly in B-decay}

While the usually dominating weak decays  are of so called charged
current type,
which have the highest rates, the first anomaly\cite{RK} called
``$R(K)$ and $R(K^*)$''
concerns in fact processes in which a pair of oppositely
charged leptons are produced, so that the lepton pair is neutral and thus
must have coupled to a neutral current, if one thinks this way.
On the figure \cite{courier} you see the diagrams in normal perturbative thinking
for the two different types of processes, charged and neutral currents.

\includegraphics{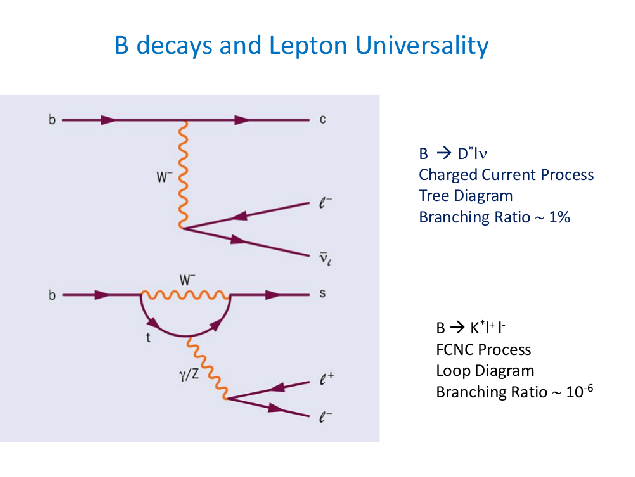}\\

The ratios studied - and found to be too small compared to the Standard
Model prediction, unity in this case - are the ratios of the decay rate of
a B-meson to respectively a $K$ or a $K^*$ and in the numerator a $\mu^+\mu^-$ pair
while in the denominator instead an $e^+e^-$ pair. One should consider the
similarity of the values measured for these  two ratios as increasing the
statistics for there being any effect at all:

\includegraphics{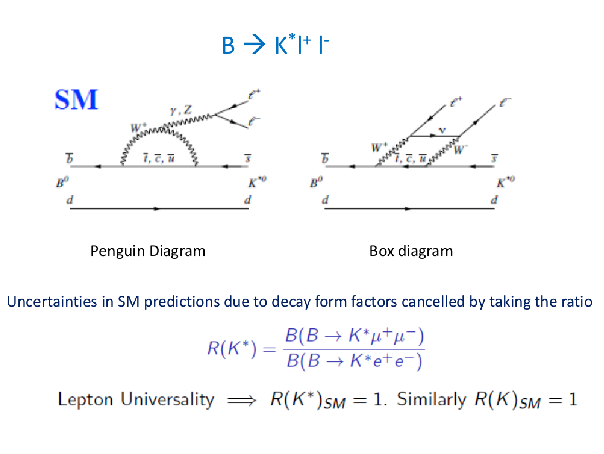}

\includegraphics{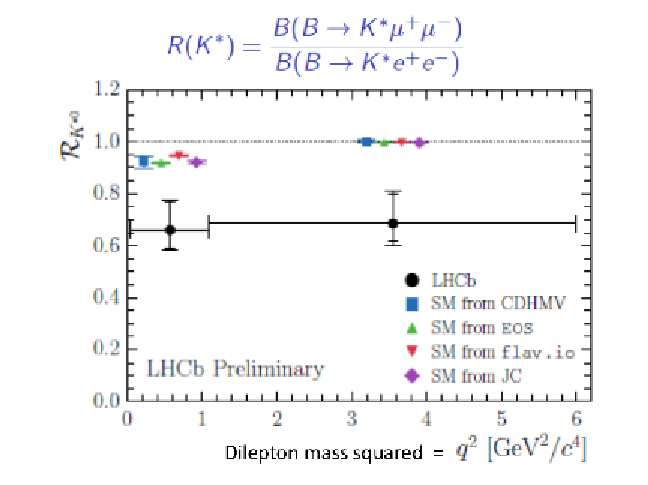}

\includegraphics{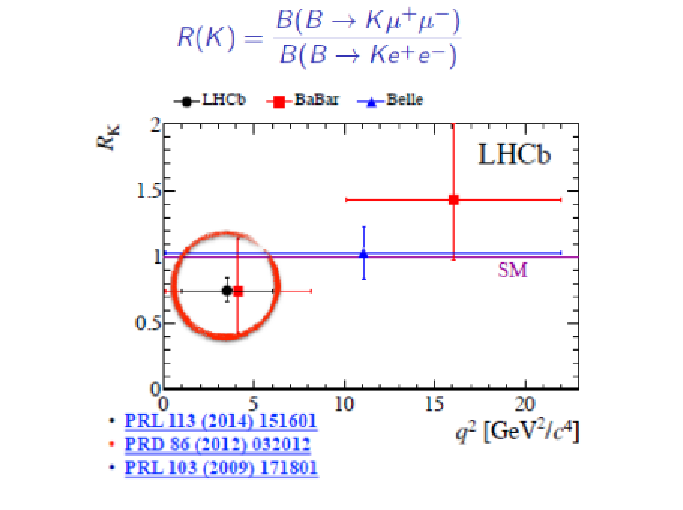}

\includegraphics{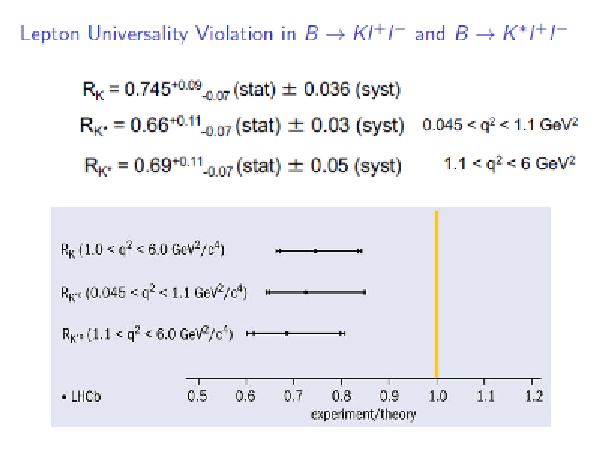}

\subsection{The $R(D)$, $R(D^*)$, and $R(J/\Psi)$ anomaly in B-decay}

An anomaly\cite{RD} has been found in the ratio
\begin{eqnarray}
R(D^{(*)}) &=& \frac{\Gamma( B\rightarrow D^{(*)}\tau \nu_{\tau})}{
\Gamma(B\rightarrow D^{(*)}\mu\nu_{\mu})}
\end{eqnarray}
(and also for the corresponding ratio $R(D)$),
which if different from its Standard Model value would represent a
lack of lepton universality between $\tau$ and $\mu$.
Measurements of decays involving electrons and muons show no deviations
with respect to the Standard Model within the current level of precision.

The Standard Model values for $R(D^*)$ and $R(D)$ differ from unity due
to the phase space difference
between the $\tau$ and $\mu$ processes:
\begin{equation}
R(D*)_{SM} = 0.252 \pm 0.003 \qquad \hbox{and} \qquad R(D)_{SM} = 0.299 \pm 0.006
\end{equation}


\includegraphics{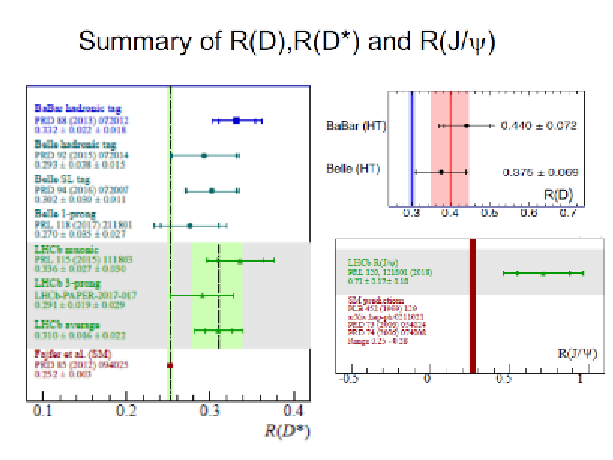}

\subsection{The $\mu$ anomalous magnetic moment anomaly}

The anomalous magnetic moments for both muon and
electron have been measured with high accuracy and compared
to the theory; only the one for the muon deviates slightly from the
Standard Model
\cite{anomalousmagn}.

\includegraphics{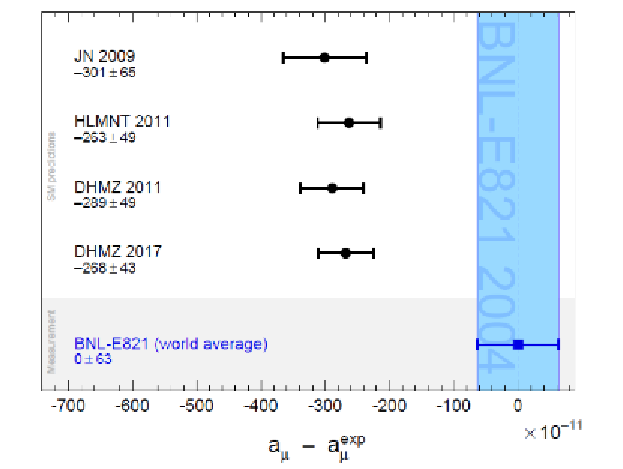}

\includegraphics{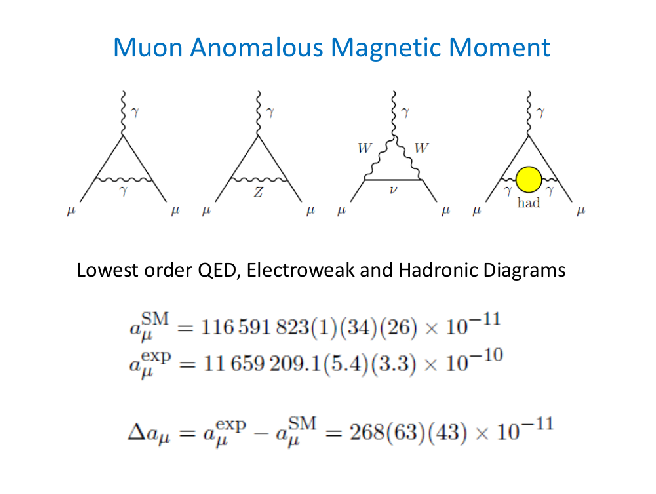}

\subsection{Proton Radius Puzzle}
In 2010 Antognini et al. \cite{Nature, RP2} measured the muonic hydrogen
 atom
spectrum so
as to find what is known as the ``Proton Radius Puzzle'', namely that the
radius
of the proton fitted to the muonic atom spectroscopy turns out $4\%$ or
5.6 $\sigma$
smaller than the radius measured in electronic atoms and e-p scattering.
A similar effect is seen in deuterium \cite{RPd}.
Using respectively electrons and muons in the atom one gets
\begin{eqnarray}
R_{p\;\hbox{ using electrons}}& =& 0.8768\pm 0.0069 fm\\
R_{p\; \hbox{using muon in the atom}}&=& 0.842±0.001 fm \quad \hbox{deviating by 5 s.d.}
\end{eqnarray}

An overview-talk \cite{pRoverview}.

\includegraphics{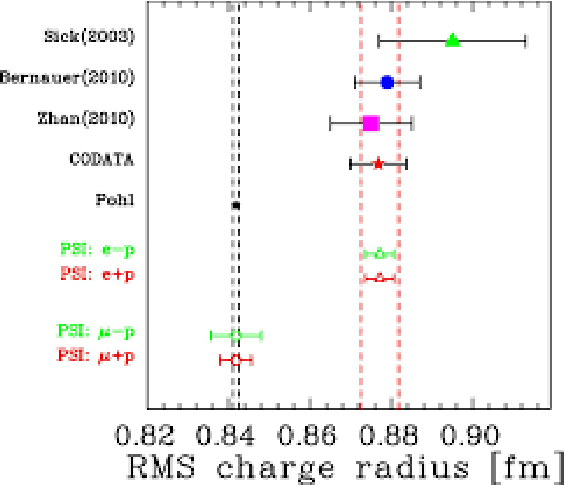}

\includegraphics{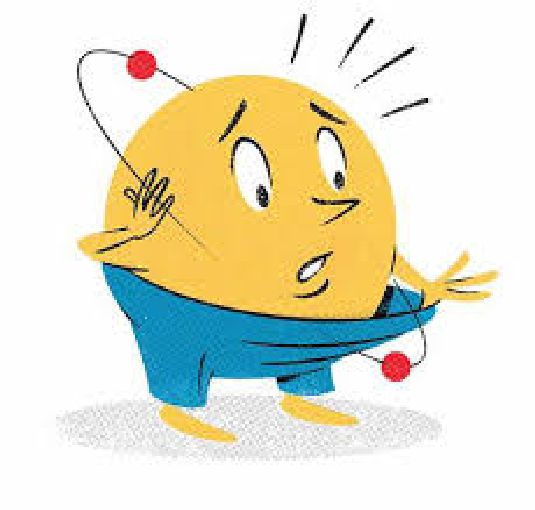}

\subsection{The $\epsilon'/\epsilon$ anomaly in $K\bar{K}$-mixing.}

\includegraphics{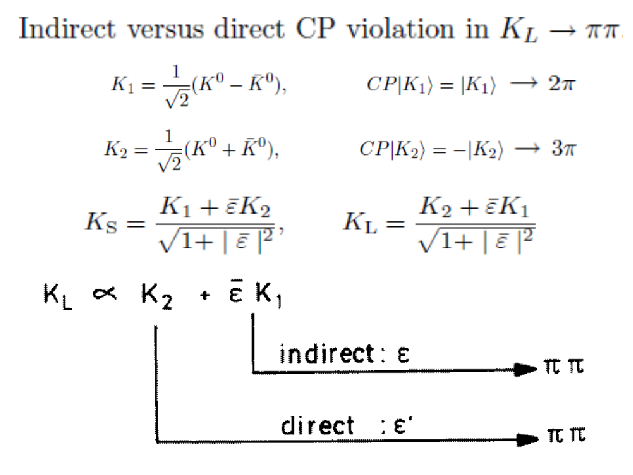}

\includegraphics{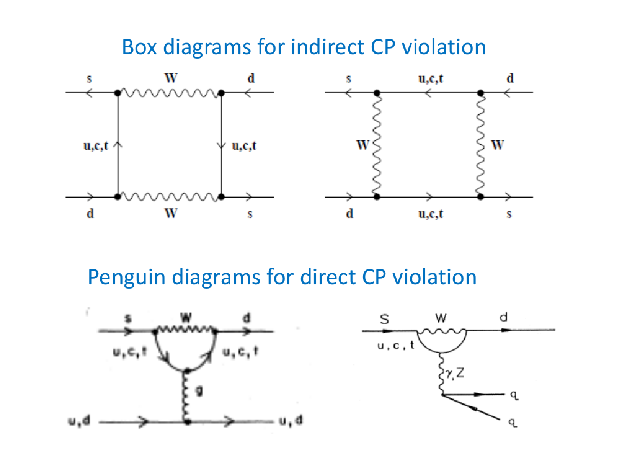}

\includegraphics{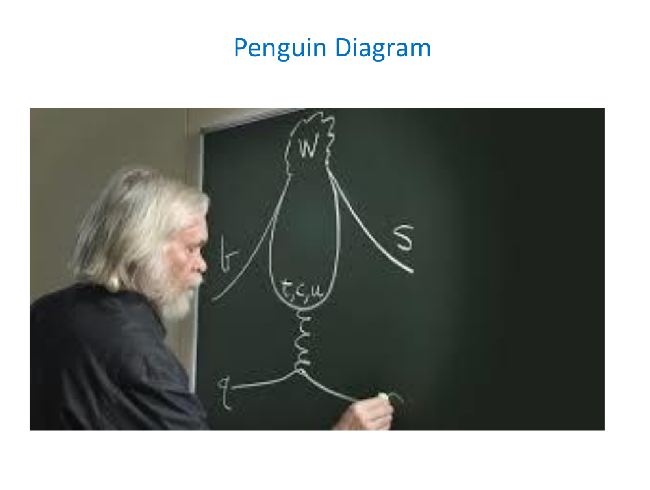}

\includegraphics{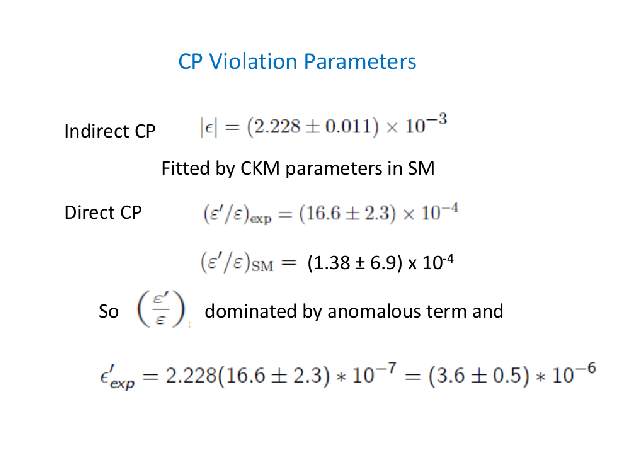}

The direct CP-violation in K --> 2 pions $\epsilon'$ is hard to calculate in
the Standard Model but gives a small value compared to the experimental
one\cite{epsilonprime, lattice} and
thus the anomaly should even dominate.

\section{Our non-perturbative model}
\label{model}

The philosophy of our model or interpretation of the anomalies is
that there is ``a new strong sector'' in addition to QCD. That is to say,
the Standard Model is actually completely correct as the fundamental model;
however, due to the appearance of the top-quark Yukawa coupling $g_t=0.935$
which is of order unity, there can and actually we suppose do appear
non-perturbative effects. We shall return below in section \ref{strong}
to the question as to what is the borderline for a coupling constant
so as to go from being one for which perturbation theory as usual
works to one for which non-perturbative effects are to be expected. We
find that, with say about 16 different particle or better components (including
color and spin variants),  the borderline coupling between strong and weak values
for the top Yukawa coupling is
\begin{equation}
g_{t \; \hbox{border}}\approx (6 \hbox{ to } 4\pi )/\sqrt{16} = 1.5
\hbox{ to  } 3.
\end{equation}
The idea is that new phenomena not found in pure perturbation
theory, such as bound states or new condensates, will show up
because of the strong coupling. We, however, seek to not commit ourselves
into details as to  what sort of new phenomena should pop up, but rather think
so abstractly about it that we only think about very high order Feynman
diagrams being important. We can then hope
our estimations are o.k. almost independently of precisely
which type of non-perturbative phenomenon pops up, bound states
or new phases,...

Indeed we assume, that the new feature in our model is some very complicated
very high order diagrams which, except for some smaller attachments to the
studied particles, consist only of vertices of the top-Yukawa coupling type.
Then of course these vertices must be connected by propagators for particles
coupling by means of this Yukawa coupling. That is to say that in the bulk
of the diagram we find only the top quarks - right and left
chirality top propagators - and the Higgs particle propagator.

\includegraphics{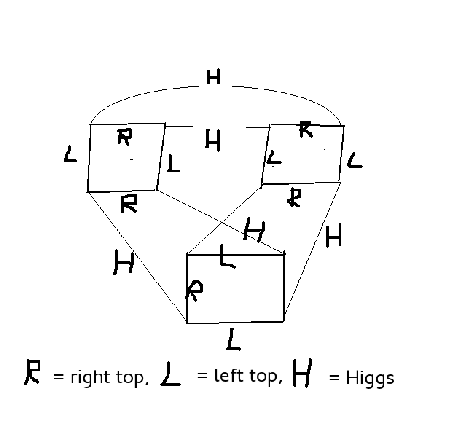}

On the figure we see a vacuum diagram with only these particle and only the
$g_t$-vertex.

\includegraphics{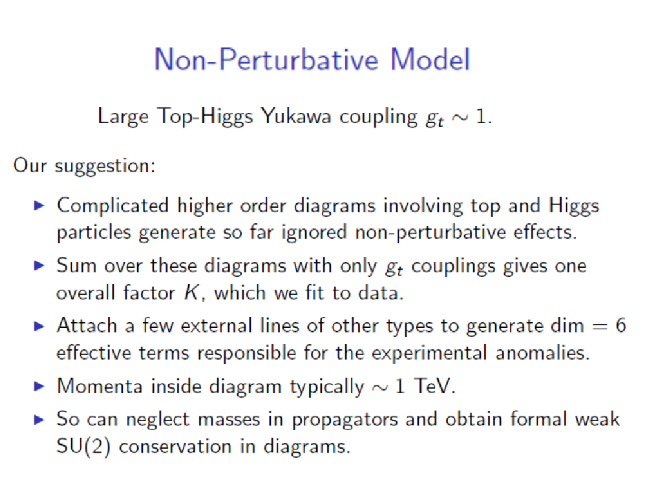}

\subsection{More precise rule}
\label{rule}
In practice we let our diagrams with a lot of $g_t$-vertices and only a few
different mainly external particles be arranged to simulate a dimension
= 6 effective field theory Lagrangian density term. This is of course a
reflection of the assumption that the whole non-perturbative effect
is short range compared to the scale of energy at which the anomalies are
observed. In fact we have in mind an energy scale of the order of
1/2 TeV, say.

To make precise sense of our overall weight factor $K$ we have to make
some rule even for the normalization of the dimension 6 field combinations
supposed to come out of our non-perturbative effect.
Using this normalization we can then say that
 the coefficient shall just be
$K$ modified by being multiplied by various suppression factors.
The rule of normalization, which we propose to use here, is that the
dimension 6 operator - to be considered properly normalized - shall be the
product of two dimension 3 operators each of which is normalized in
analogy to the expressions used to normalize the wave functions for the
particles in the following way:

Let us first remind ourselves about the way one usually normalize
wave functions:
\begin{itemize}
\item {Spin 1/2 fermion}
\begin{eqnarray}
<\psi_1|\psi_2> &=&
\int \bar{\psi}_1(x)\gamma^{0}\psi_2(x)d^3x\nonumber\\
&=&  \int \psi_1^{\dagger}(x)\psi_2(x)d^3x\nonumber\\
\hbox{or equally good}&&\hbox{ for positive energy single particle states}
\nonumber \\
<\psi_1|\psi_2> &=&
\int \bar{\psi}_1(x)\psi_2(x)d^3x\nonumber\\
\end{eqnarray}
\item{Bosons}

For Bosons with complex (non-hermitean) fields:
\begin{eqnarray}
<\phi_1|\phi_2> &=&\int  \phi_1^{\dagger}(x)
\stackrel{\leftrightarrow}{\partial_0}
\phi_2(x)d^3x
\end{eqnarray}

while for real (hermitean) fields we have a factor
2 less:

\begin{eqnarray}
<\phi_1|\phi_2> &=&\int  \phi_1^{\dagger}(x)
\partial_0
\phi_2(x)d^3x
\end{eqnarray}

The reason for this difference between the ``complex'' and the
``real'' cases, may be seen by first noticing, that if
one wants both for real and complex fields to have a propagator
of the form
\begin{eqnarray}
prop(p^{\mu}) &=& \frac{i}{p^2 -m^2}
\end{eqnarray}
 without any extra factor 2 depending on the reality, then
it is needed to take say the kinetic term in the Lagrangian
density or the Hamiltonian density to have a factor $1/2$
extra in the real case:

\begin{eqnarray}
{\cal L}(x) &=& \partial_{\rho}\phi^*\partial^{\rho}\phi +...
\hbox{ for complex field,}\nonumber\\
{\cal L}(x) &=& \frac{1}{2}\partial_{\rho}\phi\partial^{\rho}\phi +...
\hbox{ for real field.}\\
\end{eqnarray}
The point is that when you derive the propagator from the
action by functional differentiation you cannot in the
real case avoid obtaining the factor 2 from differentiating a square,
while in the complex case one can formally play as
if $\phi$ and $\phi^*$ were independent variables and avoid the
factor 2.

In fact we shall meet below, in our discussion of the anomalous
magnetic moment of the muon, a mixed case of ``complex'' with a
``real'' and propose to use as compromise a square root 2 factor,
$\sqrt{2}$.

\end{itemize}

The idea then is that we construct the dimension 6 field combination
by inserting in these expressions the second quantized field,
$\psi \rightarrow {\pmb \psi}$ and $\phi \rightarrow {\pmb \phi}$,
and multiply two of these normalized expressions together.

Next we make this a priori not relativistic expression of dimension 6
covariant, by supplementing it with components of the same four vectors
so that we get ordinary four vector contractions.
We should really even consider tensor or scalar contractions between
the two ``currents'' which are w.r.t. the 0-indices the term
going into the number of particles normalization. (In fact we shall
see that we shall use a tensor contraction for the case of the
anomaly in the anomalous magnetic moment for the muon.)

For example: After having made a product say
\begin{eqnarray}
{ \bar{\pmb\psi}}_1(x)\gamma^{0}{\pmb \psi}_2(x)
* {\bf \bar{\pmb\psi}}_3(x)\gamma^{0}
{\pmb \psi}_4(x)&&
\end{eqnarray}
we make it covariant as follows
\begin{eqnarray}
{ \bar{\pmb\psi}}_1(x)\gamma^{\mu}{\pmb \psi}_2(x)
* { \bar{\pmb\psi}}_3(x)
\gamma_{\mu}{\pmb \psi}_4(x)&&,
\end{eqnarray}
which is then our ``normalized'' dimension 6 effective term.
It should be stressed, that we must make a normalization condition like
this one in order to ever be able to have our model give more than
order of magnitude results. However, since we anyway shall
supplement by rules of inserting a factor $g_l/g_t$ for each
lepton l needed Yukawa coupling, we can hardly hope for more than
an order of magnitude accuracy anyway; but  without the normalization
specification, we would even in principle be excluded from having a
better accuracy than order of magnitude at most. At the end we find somewhat
surprisingly, that the agreement is so good, that it is almost better than
order of magnitude; then this is not a priori without meaning as
long as we use this normalization rule, but surprising anyway.

You can extract external particles without any extra - and thus lower strength -
vertices when they are the ones already participating in the big complicated
diagram. That is to say you can extract:
\begin{itemize}
\item Higgs particles both the ``radial one'' (the experimentally found
(radial) Higgs
boson)
and the ``eaten ones'' meaning truly longitudinal W's and Z's.
\item Genuine top-quarks
\item The left-handed dsb-quark combination partner of the top-quark.
This means a superposition of  three left chirality components of the
quarks {\bf d}own, {\bf s}trange, and {\bf b}ottom, which is just
that combination,
which is in the weak isospin doublet with the left chirality top quark.
These particles namely also  couple via the strong top Yukawa coupling constant $g_t$.
\end{itemize}

If we want to have other external particles in the effective coupling
to be identified as part of the non-perturbative effects, then at least
one or more extra propagators have to be added to the complicated diagram.
If the extra propagator - say a $W$ as we need in the
``The $R(D)$, $R(D^*)$, and $R(J/\Psi)$ anomaly in B-decay'' case -
is considered to sit well inside the complicated diagram, so that the
loop momenta passing through it are of the typical scale of the
non-perturbative diagram, $\sim$ 1/2 TeV, we may effectively ignore the
propagator. Then we only correct our effect by a factor coming from the deviation
of the attachment couplings of the extra propagator from the $g_t$-coupling.

The point that the ``light quarks''(meaning all but the top quark) couple only
immediately as left handed chirality particles means, that
- at least without severe extra propagators added - the most
important effective field term coming out of the model
will have its quark part solely involving left chirality
dsb-quarks (i.e. no quarks with 2/3 charge except the top itself).
Put as  a current for the quarks we thus have a left handed projector
$P_L$ inserted.

For the lepton coupling we have a priori both handednesses possible.
However, concerning the lepton, we still have to note that
it can only couple to the Higgs among the particles strongly
interacting via the $g_t$  coupling,
in as far as the leptons do not couple directly to the top quark.

Now we must further notice that on the short distance scale,
large energy scale 1/2 TeV, the vacuum Higgs field is small and is
only expected to come in, if absolutely needed. This is the
case for the magnetic moment anomaly, which is an anomaly in a
term needing its proportionality to the Higgs field expectation value
$<{\pmb \phi}>$ because the
anomalous magnetic moment coupling of the muon corresponds to a left to right
transition or opposite.

 We indeed have conservation of the weak isospin in our complicated diagrams.
So,  if an even number of the effective left chirality quarks come out of the diagram, the
number of Higgses coming out must be even too. The term we can get with say
$\bar{\pmb s}\gamma^{\nu}P_L{\pmb b}$
= $\bar{{\pmb\psi}_s}\gamma^{\nu}P_L{\pmb \psi}_b$ as the one
factor in fitting the $R(K^*)$ anomaly can have
either only left or only right say muons in the other factor, and thus be like
$\bar{\pmb\mu} \gamma_{\nu} {\pmb \mu}
=\bar{\pmb \psi}_{\mu} \gamma_{\nu}{\pmb \psi}_{\mu}$,
or such a factor with a $\gamma^5$ inserted.
We are in this way guided towards the operators $O_9$ or $O_{10}$
in the effective Hamiltonian (see below) and not even the primed operators,
 which would have signaled right handed chirality for
the quarks. Our model so to speak predicts the unprimed $O_9$ or
$O_{10}$ operators rather than the $O_9'$ or
$O_{10}'$ operators to have the anomaly.

\section{Fitting $K$}
\label{Fitting}
\subsection{Fitting $K$ for $R(K^{(*)})$ anomaly}

Indeed by fitting the effective term needed, using a single operator in addition to the pure
Standard Model, it has been found \cite{Hurth} that a value $C_9 =-1.3$ for the coefficient
to the operator $O_9$ is needed.

\includegraphics{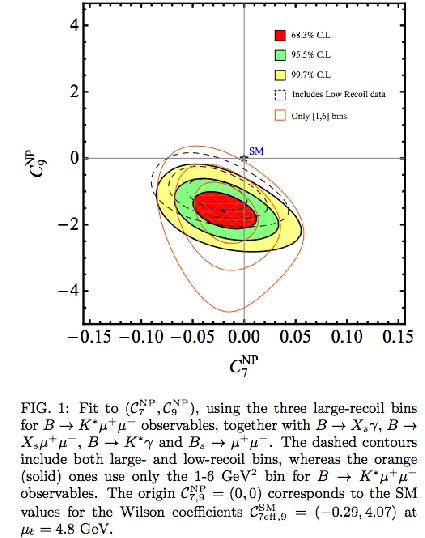}

We use this fitting to estimate the value of our overall coefficient
$K$ from the ``$R(K)$ and $R(K^*)$ anomaly in B-decay''
data:

\includegraphics{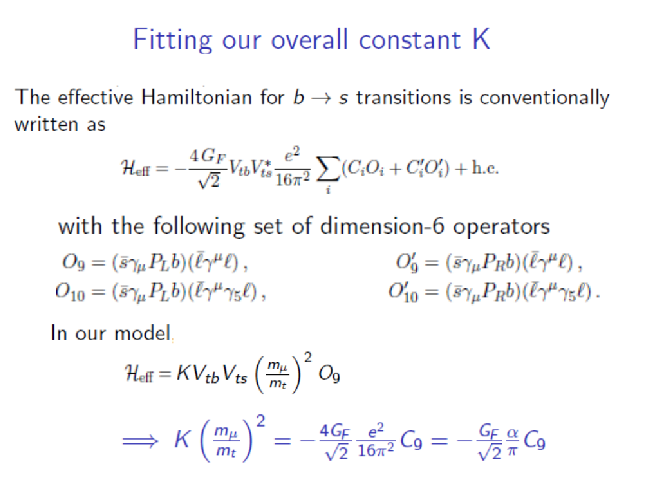}

\includegraphics{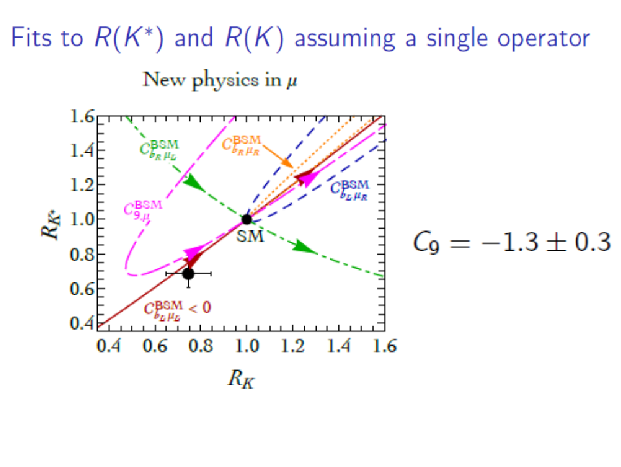}

Using
\begin{eqnarray}
G^0_F&=& 1.1664* 10^{-5}\ GeV^{-2}\nonumber \\
\alpha &=& 1/137.037...\nonumber \\
m_{\mu}&=& 105.66\ MeV/c^2\nonumber \\
m_t &=& 173 \pm 0.8\ GeV/c^2
\end{eqnarray}
we get
\begin{eqnarray}
K &\sim &- \left ( \frac{173}{0.106} \right )^2 1.1664 *10^5 GeV^{-2}*
\frac{-1.3}{\sqrt{2}\pi * 137.037}\\
&=& \frac{1}{14\ GeV^2}
\end{eqnarray}


\subsection{Fitting Our model for the $R(D)$, $R(D^*)$,
and $R(J/\Psi)$ anomaly in B-decay}

Since we do not have available any fitting of the coefficients to the
dimension 6 operators for  the $B \rightarrow D^{(*)}\tau\bar{\nu}_{\tau}$
process suspected to be the anomalous one\footnote{We indeed take, in the
cases of flavour universality violation,
the anomalous process to be the one
with the heavier one of the involved leptons in our
non-perturbative model. Thus here the $\tau$ process.},
we shall rather fit the $K$ value needed by comparing this process with
heaviest lepton one from the ``R(K) and R(K*)'' case just fitted above.

Under very crude simplifying assumptions, we indeed extract the ratio
(averaged over the phase space) of
the anomalous part for the decay
$B\rightarrow D^*\tau\bar{\nu}_{\tau}$ relative to the anomalous
part of the $B\rightarrow K^*\bar{\mu}\mu$ decay. We find that this ratio
is very close to what we expect from our model. Indeed we find, that the ratio
of the two anomaly-amplitudes is close to that of the ratio of the square of the
masses of the involved (respectively heavier) leptons.

\includegraphics{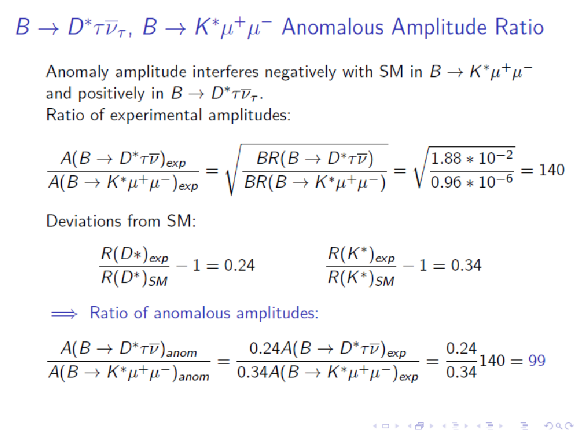}

The crude assumptions we make for our very crude approximation comparing the
two channels supposed to have an anomaly are:
\begin{itemize}
\item the amplitude is constant over all the phase space, so that it is
as if the particles only went into one single state.
\item The dominant anomalous contribution to the rate of the processes in
question is the {\em interference term } between the Standard
Model contribution and the anomalous part of the amplitude (it is the latter
part  that should be given by our estimate).
\item It is supposed that there is no anomaly in the
in the channels with the lighter one of the leptons.
So the whole anomaly in the ``$R(D^{*}$ and $R(J/\psi))$ '' case of
deviation breaking universality between $\tau$ and $\mu$ or $e$ is blamed
on the $\tau$-process. Similarly the whole deviation from lepton universality
in the ``$R(K^{(*)})$ ''' case (actually between $\mu$ and $e$) is blamed
on the muon channel.
\end{itemize}

\includegraphics
{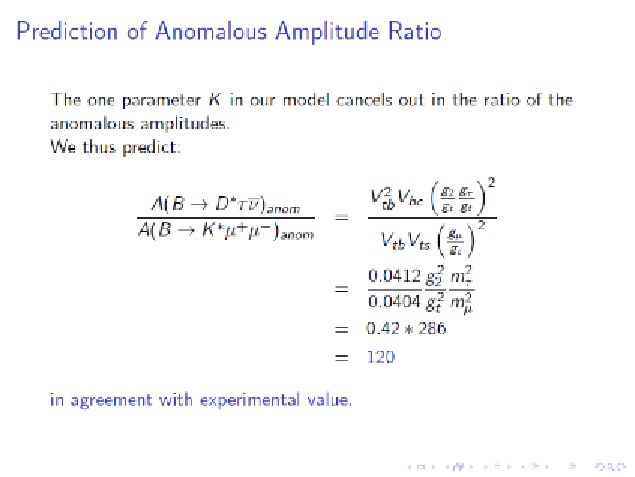}

The experimentally estimated anomaly
for the channel $B\rightarrow D^*\tau\bar{\nu}_{\tau}$ relative
to the channel $B\rightarrow K^*\mu^+\mu^-$ is 99/120 times our
prediction, which is of course actually a remarkably good
agreement. This means that, if we fitted our overall parameter $ K$
to the reaction  $B\rightarrow D^*\tau\bar{\nu}_{\tau}$, we would obtain
a $K$ value 99/120 times the value we fitted above for
the reaction  $B\rightarrow K^*\mu^+\mu^-$, namely $K =1/(14\ GeV^2)$.
Thus we can claim that the fit to the   $R(D)$, $R(D^*)$,
and $R(J/\Psi)$ anomaly in B-decay gives $K=1/(17\ GeV^2)$.

\subsection{Fitting to the muon magnetic moment anomaly}

The philosophy of our model for treatment of the nonperturbative
effects is to look at the dimension 6 effective field theory coupling
needed for providing the anomaly found experimentally, and seek to
very crudely\footnote{by changes that are at least no more than of order
unity, but preferably just looking at analogies under
Lorentz transformations} identify it with a dimension 6 operator
which is very formally of the character of a product of
two ``currents'' of dimension 3 each. Then the idea is to
normalize these two currents by a close analogy to the
normalization of the ``currents'' going into the normalizations
of the single particle wave functions relativistically.

As stated here, we must admit that the rule does not sound
extremely clear and definite. However we hope, by using the example
of the anomalous part of the anomalous magnetic
moment for the muon, to at least illustrate that there
is a hope to suggest a clear rule - if not immediately then
after a bit of experience:

Experimentally it has been found that the parameter $a_{\mu}=\frac{g_{\mu}-2}{2}$ ,
``the anomalous magnetic moment for the muon'', being half the difference
of the g-factor for the muon magnetic moment from the free
Dirac equation obtained value $g=2$, deviates by a small
amount $\Delta a_{\mu} = a_{\mu}^{exp} - a_{\mu}^{SM} = 2.68(63)(43) *10^{-9}$
from the Standard Model prediction $a_{\mu}^{SM}=116591823*10^{-11}$.
This is a priori to be explained, in the spirit of our story, by there
being for some reason or another at first a high dimension effective
field theory term
\begin{eqnarray}
\Delta {\cal H}(x) &=& \Delta a *2 \bar{\pmb\psi}(x)\vec{{\pmb B}}(x)\frac{e}{2m_{\mu}}
\vec{S}{\pmb \psi}(x)\label{magnextra}.
\end{eqnarray}
Here the ${\pmb\psi}(x)$ is the second quantized muon field and
$\vec{{\pmb B}}(x)$ is the second quantized magnetic field
${\pmb B}_1 ={\pmb F}_{23}; B_2= {\pmb F}_{31}; {\pmb B}_3={\pmb F}_{12}$. Further
$\vec{S}=\frac{1}{2}\vec{\sigma}$ is the muon spin matrix,
and e.g. $S_3 = \frac{1}{2}\gamma^1\gamma^2$. For non-relativistic particles
without spin - just moving around with mass $m$ and charge $e$ -
the ratio of the magnetic moment relative to the angular momentum
is $\frac{e}{2m}$. However when we have spin, this ratio may be
different and the correction factor is called the $g$-factor
and is here denoted by $g$.

Keeping track of the factors 2 we rewrite
the extra effective term needed to provide the anomaly
seen experimentally as
\begin{eqnarray}
\Delta {\cal H}(x) &=& \Delta a *2 \bar{\pmb\psi}(x)\vec{\pmb B}(x)
\frac{e}{2m_{\mu}}
\vec{S}{\pmb \psi}(x)\label{magnextra}\\
&=&2\Delta a *\bar{\pmb \psi}(x)\frac{1}{2}{\pmb F}_{ij}(x)
\frac{e}{2m_{\mu}}\frac{1}{2}
\gamma^i\gamma^j{\pmb \psi}(x)\nonumber \\
&=& \frac{\Delta a}{2}*\frac{e}{2m_{\mu}}{\pmb F}_{ij}(x)
\bar{\pmb \psi}(x)\gamma^i\gamma^j{\pmb \psi}(x)
\end{eqnarray}

The idea is that we should obtain such an effective term -
or rather a quite covariant form of it - let us say for the
effective Lagrangian density
\begin{eqnarray}
\Delta {\cal L}(x) &=& \Delta a *2 \bar{\pmb \psi}(x)
\frac{1}{2}{\pmb F}_{\rho\nu}(x)
\frac{e}{2m_{\mu}}
\frac{1}{2}\gamma^{\rho}\gamma^{\nu}{\pmb \psi}(x)\label{magnextra},
\end{eqnarray}
from a diagram of this type:

\includegraphics{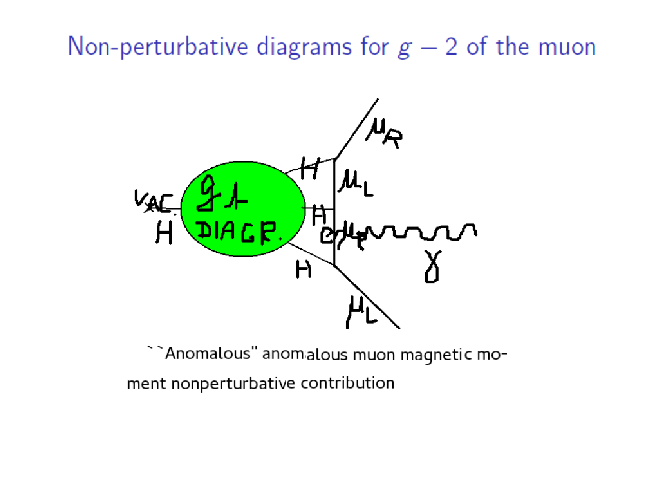}

We are led to this diagram since the anomalous magnetic moment coupling
of the muon makes a left to right transition $\mu_L \rightarrow \mu_R$,
requiring the external muon line to couple to an odd number (three) of
Higgs lines. In order to conserve weak $SU(2)$ another Higgs line must emerge
from the diagram and couple to the vacuum.

This diagram has an external leg connecting to the
vacuum Higgs field expectation value $<{\pmb \phi_H}>$, and corresponding
to this fact the terms, we asked for above, are only of dimension 5,
not of dimension 6 as we have announced we want to consider. Therefore
we shall really replace our wanted term by an equivalent dimension 6
term

\begin{eqnarray}
\Delta {\cal L_6}(x) &=& \frac{\Delta a *2}{<{\pmb \phi_H}>}
{\pmb \phi}_H \bar{\pmb \psi}(x)\frac{1}{2}{\pmb F}_{\rho\nu}(x)
\frac{e}{2m_{\mu}}
\frac{1}{2}\gamma^{\rho}\gamma^{\nu}{\pmb \psi}(x)\label{magnextra6}.
\end{eqnarray}

The exercise now is to extract from this dimension 6 term some
components, which can be written as a product of two
``currents'' of the type used to normalize wave functions
in section (\ref{rule}), so as to obtain a normalization for
the type of term we need for the anomaly
in the muon anomalous magnetic moment.

To get matching with the special role of the index $0$
in the normalization ``currents'', it may be easiest to
work with the electric field, such as ${\pmb E}_3 = {\pmb F}_{03}$ rather
than with the magnetic field we truly set out to treat. However, because
they are so closely related by Lorentz transformations, it is
not so important as far as the magnitude is concerned;
we can even keep control of the factors of 2 by such an analogy.

So let us take a Lorentz component after our
favourite technical and pedagogical choice:
\begin{eqnarray}
\Delta {\cal L_6}(x) &``contains''& \frac{\Delta a *2}{<{\pmb \phi}_H>}
{\pmb \phi}_H \bar{\pmb \psi}(x){\pmb F}_{03}(x)
\frac{e}{2m_{\mu}}
\frac{1}{2}\gamma^{0}\gamma^{3}{\pmb \psi}(x)\label{magnextra6}.
\end{eqnarray}
Note that we removed the one factor $\frac{1}{2}$ because we
now wrote explicitly indices 0 and 3 and thus do not sum
over the permuted 3 and 0.

In order to get it to match with the normalization ``current'',
which we want, we have to make the approximation of replacing the
$\gamma^3$, which at least has its square equal to unity, by unity itself.

We are looking to produce in our model the following
type of term(s), because that is what seems to have been seen by the
slight deviation of the very accurate muon magnetic
moment anomaly calculation compared to experiment:

\begin{eqnarray}
\Delta {\cal L_6}(x) &``contains''& \frac{\Delta a *2}{<{\pmb \phi}_H>}
\frac{e}{2m_{\mu}}
{\pmb \phi}_H(x){\pmb F}_{03}(x)\bar{{\pmb \psi}}(x)
\frac{1}{2}\gamma^{0}\gamma^{3}{\pmb \psi}(x)\nonumber \\
\hbox{which taking $\gamma^3 \sim 1$}&&\hbox{ gives}\nonumber\\
\Delta {\cal L_6}(x) &``contains''& \frac{\Delta a *2}{<{\pmb \phi}_H>}
\frac{e}{2m_{\mu}}
{\pmb \phi}_H(x) \partial_0 {\pmb A}_3(x) *\bar{{\pmb \psi}}(x)
\frac{1}{2}\gamma^{0}*1{\pmb \psi}(x)\label{ona3} 
\end{eqnarray}
where we have allowed ourselves to take a gauge $A_0=0$, so
that the second term in $E_3$ was neglected.

Now we have come very close indeed to writing our wanted
effective Lagrangian density term in the wanted way of
being a product of two ``currents'' with their specific normalization
attachable to the inner product normalization for wave functions.
However the boson-``current''-factor has become
{\em mixed}, in the sense that the two boson fields are from different
particles: the Higgs and the photon respectively.

A little detailed problem is that the photon field,
$A_3$ say, is ``real'' (or ${\pmb A}_3$ hermitean), while the Higgs field
$\phi_H$ is complex (or ${\pmb \phi}(x)$ non-hermitean). This difference namely
means that they should be normalized with ``currents'' deviating
from each other formally by a factor 2. Indeed, according to
the rules discussed above in section \ref{rule}, we should use:

\begin{eqnarray}
``current''|_{\psi} &=& \bar{\psi}(x)\gamma^0 ...\psi(X)\\
``current''|_{photon}&=& A_i(x) \partial_0A_i(x) \\
``current''|_{Higgs}&=& \phi^*(x) \stackrel{\leftrightarrow}{\partial_0}\phi(x),
\end{eqnarray}
where one shall have in mind that $\stackrel{\leftrightarrow}{\partial_0}$
is modulo partial integration {\em twice} the
single derivative $\partial_0$.

Let us further notice, that these ``currents'' are supposed
to be a sum over contributions from all the polarization states
for the fields being normalized: In the fermion case the
summing over the Dirac basis is understood and in the Higgs case
we imagine the contraction of the two component Higgs fields
being understood, while in our photon analogy we have
explicitly written an index $i$ supposed to be summed over,
at least over as many photon components as are relevant.

For the case of mixing a real and a complex field, we already
proposed above a {\em compromise} with a factor $\sqrt{2}$ replacing the
either 1 or 2 in the respective real and complex case for constructing
the ``current''. In fact our compromise means that we take the ``mixed''
`current'' to be modulo ``what we called partial integrations''
(i.e. if we take it that we can change sign on a derivative and let it act
to the opposite side without making any difference)
\begin{eqnarray}
``current''|_{Higgs,photon}&=& \sqrt{2} \phi^*(x)\partial_0 A_i(x)\nonumber\\
&\sim& \frac{1}{\sqrt{2}}\phi^*(x)\stackrel{\leftrightarrow}{\partial_0}
A_3(x).
\end{eqnarray}
(But now the summation index $i$ got a little bit too
formal, in as far as we cannot contract it on the other side
where we instead have the weak isospin doublet of Higgs
components. So this compromise is indeed only very formal,
but we use it anyway.)

Using this ``mixed'' ``current'', we can now write down
our wished for effective Lagrangian piece - representing
after being made covariant by extension especially also
the magnetic moment interaction, which we truly need:
\begin{eqnarray}
\Delta {\cal L_6}(x) &``contains''& \frac{\Delta a *2}{<{\pmb \phi}_H>}
\frac{e}{2m_{\mu}}
\frac{1}{\sqrt{2}}``current''|_{Higgs,photon}
\frac{1}{2}``current''|_{\psi}
\end{eqnarray}

But there is still a little problem: Above in (\ref{ona3})
we had $\phi^*\partial_0 A_3$, while the
correct $``current''|_{Higgs \; photon}$ has the summation
over an $i$ meaning over the photon polarizations.

The identification possibilities as follows:
\begin{itemize}
\item We ignore the problem of whether we have
only $A_3$ or the summation over i. Then:
\begin{eqnarray}
\phi^* \partial_0 A_3 &\sim& \frac{1}{\sqrt{2}}``current''|_{Higgs \; photon}
\nonumber\\
&\sim& \frac{1}{2} \phi^*\stackrel{\leftrightarrow}{\partial_0} A_3
\label{isum}
\end{eqnarray}
\item We count say the occurrence of the
 order of 2
i-values summed over as a factor 2 say, and thus
take
\begin{eqnarray}
\phi^* \partial_0 A_3 &\sim& \frac{1}{2\sqrt{2}}``current''|_{Higgs \; photon}\\
&\sim& \frac{1}{2} \phi^*\stackrel{\leftrightarrow}{\partial_0} A_3
\label{isumb}
\end{eqnarray}
\end{itemize}

According to our philosophy we should consider the
operator factor $``current''|_{Higgs,photon} *``current''|_{\psi}$
as normalized, so that our coefficient $K$ modified by what we
call ``suppression factors'' should in our model be the
coefficient of just this normalized term. Looking at the diagram
we propose and noting that there are three non-$g_t$ couplings, which are
muon Yukawa couplings and one electromagnetic coupling $e$, our $K$
with suppression factors becomes $K*\left (\frac{g_{\mu}}{g_t}\right )^3*e$.
I.e. our model predicts an effective term of the
appearance
\begin{eqnarray}
K*\left (\frac{g_{\mu}}{g_t}\right )^3*e
``current''|_{Higgs,photon} *``current''|_{\psi}&
&\nonumber\\
 \hbox{ to be identified with}\frac{\Delta a *2}{<\phi_H>}
\frac{e}{2m_{\mu}}
\frac{1}{\sqrt{2}}``current''|_{Higgs,photon}
\frac{1}{2}``current''|_{\psi}&&
\end{eqnarray}
This means that to get agreement for our model,
the value of $K$ must be given by the equation
\begin{eqnarray}
K*\left (\frac{g_{\mu}}{g_t}\right )^3*e&=&
\frac{\Delta a *2}{<\phi_H>}
\frac{e}{2m_{\mu}}
\frac{1}{2\sqrt{2}}
\end{eqnarray}
using the observed value of the anomaly $\Delta a = 2.68*10^{-9}$.
The derived $K$-value should of course be approximately equal to the K-values
obtained from the other anomalies.

The numerical evaluation for $K$ from this magnetic moment
anomaly for the muon gives
\begin{eqnarray}
K_{from \; magn. mom.} =K &=&
1637^3*\frac{2 *2.68 *10^{-9}}{246 GeV}*\frac{1}{2 *0.10566 GeV}*
\frac{1}{2\sqrt{2}}\\
&=& 0.16\ GeV^{-2}\\
&=&  \frac{1}{6.25\ GeV^2}
\end{eqnarray}
This calculation was made using
(\ref{isum}), i.e. ignoring the summation, while if we
instead use (\ref{isumb}) and include a correction factor
2 for the summation, we get a fitted $K$-value which is a factor of 2
smaller, namely
\begin{eqnarray}
K_{from \; magn.\;  mom.\;  w.\; summing}&=& \frac{1}{2*6.25 GeV^2}\nonumber\\
&=& \frac{1}{12.5\ GeV^2}.
\end{eqnarray}

\subsection{Explaining Proton Radius Puzzle}

We now also want to explain the  anomaly of the proton radius
fitted to the muonic atomic physics spectrum seeming to be 4\% smaller
than the fit to the usual electronic atom spectrum in our model.
The idea is of course that we shall propose some dimension 6 - like the other
anomalies were explained by dimension = 6 effective field theory terms -
effective field theory term letting the muon interact with the
hadronic matter
in the proton providing a slightly different potential for the muon, while
passing the proton.
Such a correction to the potential for the muon inside the proton may easily
be simulated by giving the proton a different radius. In fact the main
effect of diminishing the radius of the proton on the electric potential felt
by the muon - which is supposed to be the main interaction between the
proton and the muon in usual electrodynamics - is to
increase the potential a bit in the region filled by the proton.
In our anomaly model the interaction of the muon is not just
electromagnetic, as usually supposed, but partly due to an anomalous
effective field theory term letting the muon interact with
indeed gluon fields inside the proton.

The term, on which we shall settle as the one that can provide sufficiently
strong interaction between the proton and the muon, is a term of the form
\begin{eqnarray}
 {\cal L}_{eff, \mu p}&=& K\frac{g_S^2}{g_t^2} *\frac{g_{\mu}}{g_t}
 *{\pmb \phi}_H(x) {\pmb A}_{\rho}^a(x){\pmb A}^{\rho a}(x)
*\bar{\pmb \psi}_{\mu}...{\pmb \psi}_{\mu}
\end{eqnarray}
where ${\pmb \psi}_{\mu}$, ${\pmb A}_{\rho}^a$ and ${\pmb \phi}_H$ are the
fields for respectively
the muon, the gluon and the Higgs particles. The coefficient in front
of the field combination is a product of our fitting constant $K$,
the ratio $\frac{g_S^2}{g_t^2}$ of the strong gauge coupling $g_S$ of QCD
relative to the top quark
Yukawa coupling $g_t$, and the ratio $g_{\mu}/g_t $of the muon Yukawa
coupling relative to the
top yYukawa coupling. The dots "$...$" stand for some $\gamma$-matrices, but they
are really
just the unit matrix $1$ in our present case. This proposal is actually very
problematic in as far
as it is not QCD gauge invariant. One must imagine that we extend it with
further
terms in some way to make it gauge invariant. To make it properly gauge
invariant we may have to
completely replace it by some quantity of a similar nature and dimension:
\begin{eqnarray}
 A_{\rho}^aA^{\rho a} \rightarrow F_{\rho\sigma}^aF^{\rho \sigma a}/E_{gluon}^2.
\end{eqnarray}
In practice we shall be satisfied by estimating the average value of the
$AA$ factor inside the proton, just by some dimensional argument, in terms
of strong interaction scale parameters.

\subsection{Explaining our ``theory''}

The principle of our model is that we imagine that we describe a very
complicated
diagram, in which by far the most copious vertex is the $g_t$ proportional
vertex
coupling the top quark or its doublet partner to the Higgs doublet. Then one
associates the particles present in the effective field theory term we seek
to obtain with the diagram,
by slight modifications of such an otherwise only $g_t$ coupling diagram. Since we
need a muon involved,
we must at least extract one Higgs propagator and attach it to the muon by
the Higgs muon Yukawa coupling $g_{\mu}$. This coupling couples a right
chirality muon to
a left chirality one. In addition we must have the gluon fields attached, by
the gluons
coupling to the color of some top quarks in the complicated diagram with
only the $g_t$ coupling.
This gives rise to the appearance of the extra factor $g_S/g_t$ one for each
extracted gluon (field).
By having two gluon fields extracted we can at least achieve gauge
invariance under
the global (part of the) gauge symmetry. Remembering that our diagram with
only $g_t$ coupling
and propagators taken to have no masses - and the propagators at all
being basically ignored -
conserves the
weak isospin $SU(2)$, we must have one more Higgs extracted, since otherwise
the scalarly
coupled muon would take off weak isospin. Therefore we had to have the
extra field
$\phi_H$ in the effective field theory term. We shall dispose of that by
inserting for it its vacuum value $<{\pmb \phi}_H> = 246 GeV$. Notice that we
managed to
make the term of dimension = 6, as in the other anomalous terms we proposed.

With non-relativistic muons the Dirac field combination
$\bar{\psi}_{\mu}\psi_{\mu}$ just
becomes the numerical square of the non-relativistic muon field in the case
of a
single muon, and thus gives the muon (probability) density. Calculating
perturbatively the important correction to the muon Schr{\o}dinger equation from
our anomaly now becomes a correction to the potential $V_{anomaly}$ felt by the
muon while being present inside the proton. This correction is given as
\begin{eqnarray}
 V_{anomaly} &\sim& K\frac{g_S^2}{g_t^2} *\frac{g_{\mu}}{g_t}
 *{\pmb \phi}_H {\pmb A}_{\rho}^a{\pmb A}^{\rho a}
\end{eqnarray}
or rather we should estimate the average value of this operator
inside the proton to obtain the potential it provides there.

The expectation value of ${\pmb \phi}_H$ is of course easily
taken to be $<{\pmb \phi}_H > = 246 GeV $ as is well-known in vacuum.

From dimensional arguments the $<{\pmb A}{\pmb A}>$, although not meaningful
for gauge reasons, may be taken to be an energy squared at the strong interaction
scale such as $\Lambda_{ QCD}^2 = (220 MeV)^2 = 0.05 GeV^2$.

We could also take the bag constant $B = 10^{-3} GeV^4$ from the
MIT bag model and divide it by the needed square of a supposed
energy of a gluon.
Now it is possible to calculate the lowest
energy level for a massless particle in a sphere of radius $R_p$
and that is found to be $\omega_1/R_p$ where $\omega_1 = 2.04$.
Using $R_p =5 GeV^{-1}$, we obtain
$E_{gluon} \sim 2.04/5$ GeV $\sim 0.4$ GeV.
In this way
we estimate for a dimension $[GeV^2]$ quantity extracted from
QCD
\begin{eqnarray}
 \frac{B}{E_{gluon}^2} &\sim & \frac{10^{-3}\ GeV^4}{0.16\ GeV^2} = 0.006\ GeV^2
\end{eqnarray}
 deviating by a factor 8  from the QCD-lambda estimate, but the
QCD-lambda is rather arbitrary.


Then, taking $K \sim \frac{1}{15\ GeV^2}$,
the anomalous potential felt by the muon inside the proton in our
model becomes
\begin{eqnarray}
 V_{anomaly} &\sim& K* \frac{g_{\mu}}{g_t} 246\ GeV * 0.006\ GeV^2\\
&\sim & K* \frac{1}{1700} 1.5\ GeV^3\\
&\sim& 6 *10^{-5}\ GeV
\end{eqnarray}

\subsection{The potential observed $V_{exp, \mu}$}
Suppose we interpret the seemingly shrinking
proton radius, when investigated by muonic atomic studies rather
than by an ordinary electronic atom, as a change in the
potential felt by the muon in the inside the
proton. Then, if we treat the proton as a conducting object to evaluate
the electric potential, the 4 \% decrease in radius in going from
electron to muon means that the potential in the interior of the proton
goes as follows:
\begin{eqnarray}
 \frac{\alpha}{R_p}&\rightarrow \frac{\alpha}{R_p - 4\% *R_p}\\
\hbox{meaning that}      &&\\
 V_{exp, \mu}- V_{from  \; electron} = \Delta V &\approx & +4\% *\frac{\alpha}{R_p}\\
&=& 4\% \frac{1}{137 * 5\ GeV^{-1}}\\
&=& 6*10^{-5}\ GeV
\end{eqnarray}

\subsection{Conclusion for Proton Radius Puzzle}

The accidentally perfect agreement for $K=\frac{1}{15 GeV}$ means that
we would fit $K$ to this value, if we use the proton radius puzzle
anomaly to fit it.

If our picture is right - what our good agreement of course indicates -
then the physically true radius of the proton is the one determined
by the electronic atoms, for which there is a negligible anomaly. The physical
effect, when one uses muons instead, is that there is an anomalous
interaction between the muon and the hadronic matter in the proton.
This interaction is attractive and functions as if the electric
potential for the muon inside the proton material were deeper, but such
an effect of a deeper potential for the negatively charged muon
can be simulated by the electric potential inside the proton material
being higher, an effect that could have been achieved by lowering
the proton radius so that the positive charge of the proton were concentrated
closer to the center of the proton. So it is a simulation effect
and the true radius is the one gotten by means of the electron.

\subsection{Direct CP-violation $\epsilon'$}

In our analysis we assume that a $\Delta I = 1/2$ rule applies to our
non-perturbative CP violating $K \rightarrow 2\pi$ decay amplitudes similar to that for
the dominant CP conserving Standard Model amplitudes $Re A_0$ and
$Re A_2$. So, in the notation of
Winstein and Wolfenstein \cite{Winstein}, we take
\begin{equation}
\frac{Im A_2^{anom}}{Re A_2} \sim \frac{Im A_0^{anom}}{Re A_0}
\end{equation}
Thus our non-perturbative $I = 0$ and $I = 2$ decay amplitudes should give
similar contributions to $\epsilon'$. So, for simplicity, we just treat
our non-perturbative $K \rightarrow 2\pi$ as having isospin $I = 0$ to
estimate the value of $\epsilon'$.

\includegraphics{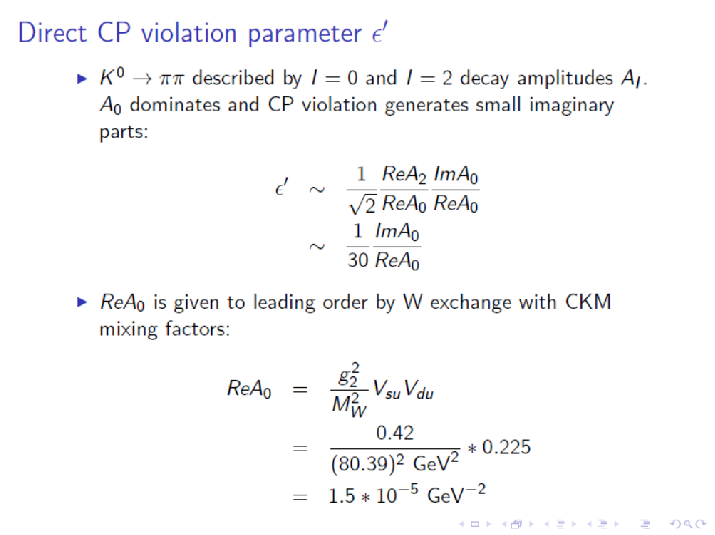}

\includegraphics{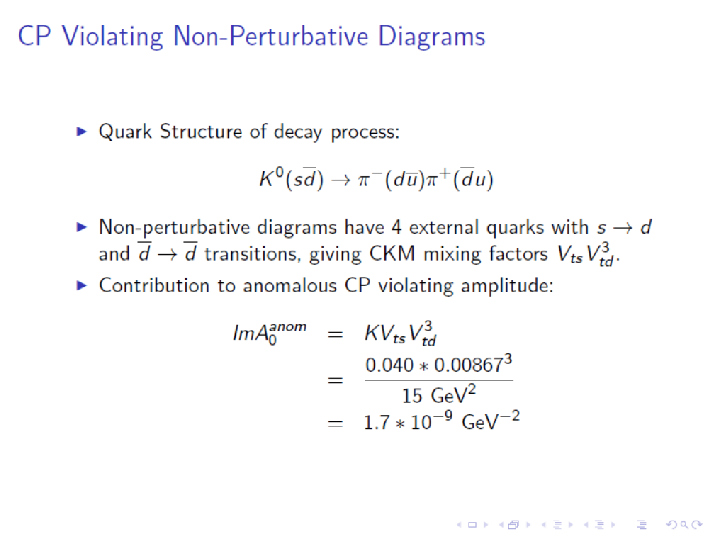}

\includegraphics{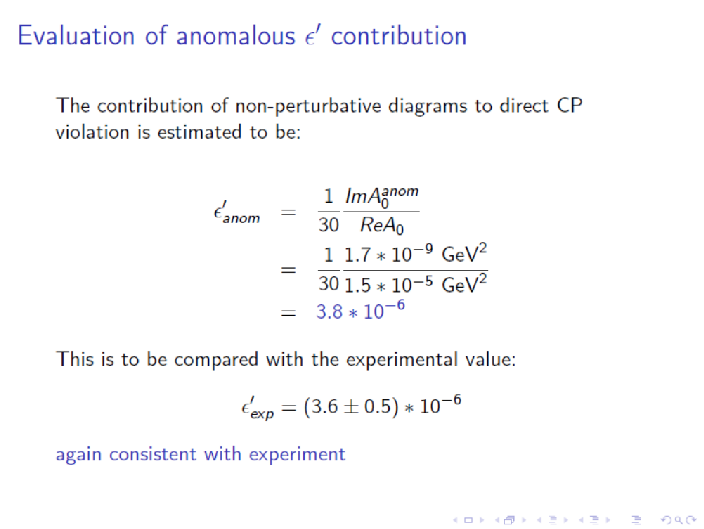}

Remember that the Standard Model prediction for $\epsilon'$ is relatively
small, and thus $\epsilon'_{exp}$ can roughly be identified with the anomaly
- especially if the anomalous and Standard Model contributions
have relative random phases. So that when we
get a factor $\frac{3.8*10^{-6}}{3.6*10^{-6}}$ too big an
anomaly, having used that $K$ were $\frac{1}{15 GeV}$, it means that a fit
of our model to
$\epsilon'$ would give $K = \frac{1}{16 GeV^2}$.

\section{R\'esum\'e Table}

The remarkable point of our whole investigation is that the different
values of the overall scale parameter $K$ fitted to the different anomalies,
indeed very different types of experiments, gets at least order of magnitudewise
very much the same value from all the anomalies considered. This is seen
in the following table:

\label{resumetable}
\begin{tabular}{|c|c|c|}
\hline
Observable & Fitted $K$ &\\
\hline
$B \rightarrow K^* \mu^+\mu^-$ & $\frac{1}{14} GeV^{-2}$&\\
$B\rightarrow D^* \tau \nu_{\tau}$&$ \frac{1}{17} GeV^{-2}$&\\
$a_{\mu}$ & $\frac{1}{12}GeV^{-2}$ &\\
$\epsilon'$& $\frac{1}{16} GeV^{-2}$&\\
Proton radius& $\frac{1}{15} GeV^{-2}$ &\\
\hline
\end{tabular}

Actually note that
the deviations between the
different fitted values of $K$ are only of the order of 20 \%. But our calculations are typically
so crude that agreement of the order of 20\% must almost be considered only
an accident. At least it would require some further argumentation to justify
that we could do it with such an accuracy. Such argumentation
could  e.g. be saying that in the case of the two $B$-decay anomalies
indeed a so small part of phase space plays the main role that a constant
amplitude over the whole phase space would be justified.


\section{Strong}
\label{strong}

The  thesis of the talk suggested from the
$5-1 = 4$  ratios of anomaly strengths agreeing, because of
of fitting 5 anomalies with the
same $K$ value is:
{\bf The  Standard Model is  perfectly O.K.
even with Anomalies,
provided
one includes non-perturbative effects}
not only from Q.C.D., {\bf  but
also from a strong sector due to
the Top Yukawa coupling $g_t$
being ``strong''.}

 Does this now  mean: No New Physics ?
{\color{blue} Logically :}
{\bf Yes, no new physics!}
but
{\color{blue}in reality:}
The explanation for why $g_t$ is so strong?
is suggested to be a new principle of ours,
the so-called
{\bf Multiple Point Principle \cite{MPP}:}
 Nature likes the couplings, such as
e.g. $g_t$, to be critical, i.e. on a phase
border (of some phase transition say)!

We and others have stated the
``multiple
point principle'' in the form: the coupling constants
get fine tuned so as to make several vacua
have the same/or very small energy
density.

In any case we know from the studies of supernovae etc.
in astronomy, that the energy density (= cosmological constant)
is very very small compared to any contribution to the
energy density that could be expected from various sub-theories of the Standard
Model, such as the strong interactions, or electrodynamics,.... Thus,
from the particle physics point of view, the energy density of the vacuum is
- in fact very mysteriously - extremely small. Thus postulating many
vacua to exist, which have the same energy density would anyway imply
that all the vacua have almost zero energy density. We thank Leonard Susskind
for the proposal to formulate the multiple point principle as several vacua
having $\sim$ zero energy density. By doing that you absorb the
mysterious almost zero cosmological constant into the postulation
of the multiple point principle.

\subsection{ Analogy and Deviation from Q.C.D.}

{\bf Analogies:}
At the scales we can care for, the ``weak '' and
``strong '' scales,
experimentally:
\begin{itemize}
\item Both
\begin{eqnarray}
\sqrt{\frac{8 +18?}{4\pi}}g_S&=&
\sqrt{(8+18?)\alpha_S}
\approx 1\nonumber\\
\hbox{ and}&&\nonumber \\
\sqrt{\frac{ 16?}{4\pi }}*g_t&\approx& 1,
\nonumber
\end{eqnarray}

(see the significance of just this size below in subsection
\ref{cv}.)
\item and they both run stronger towards
the low energy scale ($t--> -\infty$) and
weaker towards the high energy scale ($
t --> +\infty$)
(asymptotic freedom, almost for $g_t$
 too):
\begin{eqnarray}
\frac{dg_t(t)}{dt}=\beta_{g_t} > 0&\hbox{and}&
\frac{d\alpha_S(t)}{dt}=\beta_{\alpha_S} > 0 \nonumber
\end{eqnarray}
\end{itemize}

{\bf Difference between $g_t$ and $g_S$:}

The scale for {\bf $g_t$} being ``strong''
seems to be {\bf connected} to the Higgs scale
??

But the {\bf Q.C.D. scale} seems to be {\bf not
connected}
with any Higgs-like field scale.


Our multiple point principle may give
the explanation, that the Higgs field
scale is {\bf adjusted to
(fine tuned by our principle!) the
scale of
the
strong scale $t$, where $g_t(t)
\approx 1$}, because the strong $g_t(t)$
produces a new phase ``at the scale $t$''.
In fact we have previously estimated \cite{coincidence} the value
of the top quark Yukawa coupling at the phase transition to be
\begin{equation}
g_t|_{phase\ transition} = 1.00 \pm 0.14.
\end{equation}

\subsection{How strong is strong?}
\label{cv}
 Except for $\alpha_S$, the strongest coupling in the
Standard Model is the top quark Yukawa coupling $g_t$.

Let us look theoretically for how strong/large a coupling
shall be to cause non-pertubative effects by Feynman diagrams with
very many  vertices and propagators becoming important, i.e. not becoming
small.

We may crudely perform such an investigation by imagining going from one diagram
- let us say already with many vertices and propagators - to one
with one loop more. Adding one loop means, e.g. for
Feynman diagrams like the ones with only $g_t$-vertices having three
propagators attached, that one adds three propagators and two vertices.
And then of course there shall be
one more $d$-dimensional integral
in the momentum representation, where $d$ is the dimension of space-time.
In the case of the $g_t$ diagrams with dimensionless coupling
constants $[g_t] = 1$, the extra integration will for
large loop momenta get just equally many extra denominators as there are
extra integrations in the $\frac{d^dq}{(2\pi)^d}$ over the loop momentum
$q$. In the large loop variable approximation we thus add, for each extra
loop, a formally logarithmically divergent factor of the form
\begin{eqnarray}
g_t^2\int \frac{O(1)}{(q^2+...)^{4/2}}\frac{d^4q}{(2\pi)^4}\label{extraf}
\end{eqnarray}
for $d=4$. When this extra factor is found inside a convergent
diagram or if the divergencies are taken care of somehow by renormalisation,
we may not have to take the logarithmic divergence so seriously, but
rather take it as roughly unity.

\includegraphics{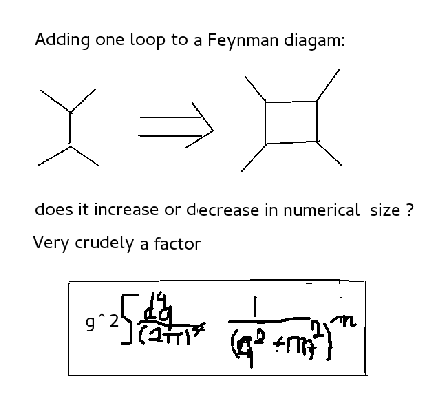}

Now, in order to investigate what is the border line value between strong and weak
coupling, we shall estimate when this extra factor is bigger or smaller
than unity, because that signals whether the diagram of higher order
- more loops - gets bigger or smaller than the foregoing one in the
construction series. So the crudest estimate of the borderline coupling size
is to require the coefficient to the logarithm in the logarithmic
divergence appearing in   (\ref{extraf}) to be unity.


{\bf The Borderline Coupling between Weak and Strong for
Only One Component is $g \sim 4\pi$.}

Taking very crudely by a ``dimensional
argument''
\begin{eqnarray}
\int \frac{d|q|}{|q|}& \sim&  1
\hbox{  (by dimensional argument)}
\nonumber\\
 \hbox{and the borderline}&&\hbox{coupling $g_{border}$ to}
\nonumber\\
\hbox{ have
the extra factor by adding}&&\hbox{ a loop to satisfy}\nonumber\\
g^2\int \frac{d^4q}{(2\pi)^4 |q|^4} &\approx& 1
\hbox{  (ignoring
the mass squares}\nonumber\\
\hbox{ in the propagators)}&&\nonumber\\
\hbox{So we get}&&\nonumber\\
g_{\hbox{border}} &\approx &
\sqrt{\frac{(2\pi)^4}{\pi^2}} =
4\pi.
\nonumber
\end{eqnarray}

{\bf Rydberg constant $\approx$ Mass for a
 Coupling of ``Order Unity''}
\begin{eqnarray}
 R_{\infty }&=&{\frac {\alpha ^{2}m_{\hbox{e}}c}{4\pi \hbar }}
={\frac {\alpha ^{2}}{2\lambda_{\hbox{e}}}}
={\frac {\alpha }{4\pi a_{0}}}\nonumber
\end{eqnarray}
is of the {\bf order of the mass-energy
$m_{\hbox{e}}c^2$} for
\begin{eqnarray}
R_{\infty} &=& m_{\hbox{e}} c^2\\
\hbox{implying}&&\\
1&=&  {\frac {\alpha^{2}}{4\pi c \hbar }}\\
\hbox{or}&&\\
\alpha^2&=&4\pi \hbox{   for $c=\hbar =1$.}\\
\hbox{meaning}&&\\
 \hbox{e} &=&\sqrt[4]{(4\pi)^3}\\
&=& (4\pi)^{3/4} \approx 6
\end{eqnarray}
{\bf Size of Coupling and Number of ``Components''}

If there were e.g. a color quantum number
taking
{\bf $N$ values for the particle types}
encircling the loop,
then there would be $N$ various loops for each one in
case of no such inner degree of freedom. According
to our philosophy of the extra factor by inserting a
loop being of order unity
\begin{eqnarray}
g_{\hbox{border}}^2N\int \frac{d^4q}{(2\pi)^4 |q|^4} &\approx& 1
\end{eqnarray}
then the $N$-dependence of the borderline coupling
between the perturbative and non-perturbative regimes would be
\begin{equation}
g_{\hbox{border}} \propto \sqrt{\frac{1}{N}}.
\end{equation}
{\bf For say 16 ``Components'' Borderline
Coupling $\sim$
1.5 to 3}

Very crudely counting particle and antiparticle also as
different ``components'' and counting together both the
Higgs with its 4 real components and the top quark with
its 3*2*2=12 components, we get in total for the particles
interacting via the top Yukawa coupling $g_t$ just 12+4 = 16
components. Thus the borderline value for $g_t$ becomes
\begin{equation}
g_{t \; \hbox{border}}\approx (6 \hbox{ to } 4\pi )/\sqrt{16} = 1.5
\hbox{ to  } 3.
\end{equation}
Experimentally
\begin{equation}
g_{t \; \hbox{exp}}= 0.93_5
\end{equation}


\section{Neutral meson mixing, $K⁰$ $\bar{K⁰}$ mixing.}
\label{mixing}
Our model for the anomalies, which seems so successful so far,
has the problem that it predicts a far too big anomaly for e.g.
$K^0$ $\bar{K^0}$ mixing. Lattice calculations of this mixing
and the mass difference of the two CP-eigenstates, or almost equivalently of the
$K_S$ to $K_L$ mass difference, provide a Standard Model prediction \cite{Bai}
of
\begin{eqnarray}
\hbox{Bai (thesis):}
\Delta M &=& (5.5 \pm 1.7)*10^{-12}\ MeV\label{Bai}\\
\hbox{ Experiment:} \Delta M &=& (3.483 \pm 0.006 )*10^{-12}\ MeV\label{Experiment}
\end{eqnarray}
Although there would here be place for a negatively interfering anomaly
of the same order of magnitude as the Standard Model prediction, there would of
course not be place for a $3*10^4$ times larger anomaly than the Standard Model prediction.
Taking our model straight away would give something of that order, so in this
sense our simplest model is falsified. However such a relatively huge effect
of our model pops up precisely in these mixing processes, because the amplitude
or coefficient of the dimension 6 effective field term is only suppressed from our
surprisingly large $K$ value by some
mixing matrix element
factors. The suppression is ``only'' by a factor $V_{td}V_{td}^* V_{ts}V_{ts}^*$
and thus the effective Lagrangian density term becomes of the order
\begin{eqnarray}
 &&K*V_{td}V_{td}^* V_{ts}V_{ts}^* \bar{\pmb \psi_s}...{\pmb \psi}_d
* \bar{\pmb \psi}_s ...{\pmb \psi}_d\\
&=&
\frac{1}{15 GeV^2} * |0.0404|^2 *|0.00867|^2 \bar{\pmb \psi_s}...{\pmb \psi}_d * \bar{\pmb \psi}_s ...{\pmb \psi}_d\\
&=& \frac{1}{15 GeV^2}*1.2*10^{-7}\bar{\pmb \psi_s}...{\pmb \psi}_d
* \bar{\pmb \psi}_s ...{\pmb \psi}_d
\end{eqnarray}
while some of our other anomalies involve in addition Yukawa couplings for
$\mu$ or
$\tau$ leptons. Our dots "$...$" could typically be $\gamma_{\mu}$ and in fact it is likely to be
accompanied with
a $\gamma_5$ projection say.

Our effective field theory term here should be compared to the effective term
estimated from the Standard Model by M. Gaillard and B. Lee \cite{Gaillard},
 which
gives the correct order of magnitude for the mass difference
\begin{eqnarray}
 {\cal L}_{eff}&=& -\frac{G_F}{\sqrt{2}}\frac{\alpha}{4\pi}\epsilon_0
\cos^2\theta_C\sin^2\theta_C
\bar{\pmb \psi_s}\frac{1}{2}(1-\gamma_5){\pmb \psi}_d
* \bar{\pmb \psi}_s\frac{1}{2}(1-\gamma_5){\pmb \psi}_d
\end{eqnarray}
where
\begin{eqnarray}
 \epsilon_0 &=& \frac{\delta}{m_W^2\sin^2\theta_W}\\
 &=& \frac{\delta}{(38\ GeV)^2} = \frac{m_c^2}{(38\ GeV)^2}.
\end{eqnarray}
with a charm quark mass  $m_c = 1.3$ GeV.
So the ratio of our anomalous coefficient compared to the Standard Model
coefficient, which fits experiment rather well, is
\begin{eqnarray}
 \frac{\hbox{ ``Our coefficient''}}{\hbox{``S.M. coefficient''}} &=&
\frac{ \frac{1}{15\ GeV^2}*1.2*10^{-7}}
{-\frac{G_F}{\sqrt{2}}\frac{\alpha}{4\pi}\epsilon_0\cos^2{\theta_C}
*\sin^2{\theta_C}}\\
&=& \frac{8*10^{-9}}{m_c^2 * 1.6 *10^{-13}\ GeV^{-2}}\\
&=& 5/(1.3)^2 *10^4 = 3*10^4.
\end{eqnarray}

Speculating that the typical energy scale at which our anomaly
terms set in is of the order of $\mu = 1\ TeV$ or the corresponding distance
0.2 am ( $= 0.2*10^{-18}$ m), our effective coupling of
$8*10^{-9}GeV^{-2}$ is in the units of this scale numerically
\begin{eqnarray}
 \frac{8*10^{-9}\ GeV^{-2}}{TeV^{-2}} &=& 10^{-2}.
\end{eqnarray}
That would be o.k, but a priori one could have wondered if
it would have been an unbelievably strong coupling.

However, if we think in terms of eigenstates of the conserved quantum numbers
for our sub-theory with only the $g_t$ coupling present, the particle to consider
is say the dsb-quark linear superposition being in a weak doublet with the top
quark. The whole such linear combination will not have these various
mixing angle suppression factors and the effective field theory
coefficient would rather be just $K$, so that we would wonder if the
effective coupling in this language is possibly enormously strong.
Again taking it that the scale is given by $\mu = 1 TeV$ would now have
\begin{eqnarray}
 \frac{K}{TeV^{-2}} &\sim & 10^4,
\end{eqnarray}
and the coupling of these eigenstate particles would be
unbelievably strong. Really it is presumably not possible to
have couplings that are much bigger than unity in terms of the scale as
unit.

In our model, wherein we have non-pertubative interactions, we
may think of what happens at the micro-level as being the appearance of big
bound states composed of many constituents. Then the absurdity of
overly strong couplings would physically show up by the interacting
particles hardly being able to penetrate into each other during the
interaction.

 Thus we shall now investigate in detail the possibility for
the eigen-superpositions of quarks etc. penetrating into each other.
We hope to argue that after all there will be difficulties in the
particles penetrating into each other, and that thus the a priori
effective coupling predicted in our model, relevant for the mixing of
$K^0$ $\bar{K^0}$  and forming the mass difference of the
two superpositions $K_L$ and $K_S$, will turn out not to be so large
in reality.

\section{Penetration Effect}
\label{Penetration}
On a scale of length smaller than the scale $\mu\sim 1TeV$ at which
the effective field theory is supposed to be used, we typically
will have to think of the particles interacting in the effective field theory
as being maybe even composite or at least consisting of a
blob of surrounding constituents of virtual particles. In our scheme
aiming at taking into account the effect of our ``new strong interaction''
of the particles interacting with the top-yukawa-coupling $g_t$, the
important blob will mainly consist of three types of particles: the
right top quark, the left top quark doublet and the Higgs, including
the left dsb-quark as part of left top quark doublet.

When such blobs scatter in the way to be described by the effective
field theory, it must then mean that  these blobs are penetrating into each
other while
their constituents interact. If the interaction is sufficiently weak
the blobs will, in first approximation, penetrate deeply through each other
and the interaction may be thought of as a perturbative interaction between
the constituents of the blob around one external
particle with the blob around another external particle. However if the
interactions are
strong, one risks that one external particle together with its blob
gets absorbed before even reaching very far into the other blob.

In principle this consideration gives us the opportunity to define
a penetration depth for each species of external particle, telling
how deep  that sort of particle will penetrate before getting
totally absorbed or transformed into a quite new particle.

In order to treat the calculation of $\epsilon'$
we shall find it
useful to work with eigenstates of the CP-operator (in a certain frame).

\subsection{Eigenstates of penetration interaction}

In order to be able to describe relatively simply the absorption rate
of various particles penetrating or attempting to penetrate
into blob(s) around the other (external) particles, it is convenient to
introduce some eigenstates of the absorption rate for these particles.

We should, so to speak, diagonalize the external
particle species matrix of absorption or penetration degree.

In the first approximation we have our picture that the blob
through which the particles with their blobs have to pass, consists of
the particles/constituents, which interact via the Yukawa coupling $g_t$.

Naively one might therefore think of the medium, caused by the other
particles than the one considered, could be taken as being quite independent
of these other 3 particles; but this is for our treatment - especially our
study of the the $\epsilon'$ CP-violating process -
not a good idea.
Rather we shall take it that the blob or medium, caused by the particles
other than the considered particle, can be so to speak in two different
states: one with CP = + and one with CP = -. We shall then imagine that
this CP-value for the medium is felt all over the medium. That is to say the particle
passing through the medium will be absorbed in a way depending on
this CP-eigenvalue, wherever in the medium it happens to get absorbed.

Thus, in order to specify the absorption rate for various particles, one has to
have  in mind that we have to discuss it for each CP-eigenvalue
of the medium separately. Of course there is a very strong analogy
between the two CP-eigenvalues for the medium: The absorption is
strong whenever the quark-antiquark superposition (see discussion below) has the
same CP-eigenvalue as the medium and weak when the eigenvalues
of the medium and quark-antiquark superposition are different.

Much of the needed  diagonalization is rather trivial to perform -
at least in the approximation that we ignore the interaction with
the Higgs  vacuum expectation value and thus the different masses
for the quarks and leptons - since it is only the specific linear combination
of the three dsb-quarks, which form the weak doublet partner of
the left chirality top-quark, that can interact via the
$g_t$-coupling with the (eaten) Higgs bosons. In addition to these particles, it is only the
top quark itself that interacts via the $g_t$-coupling. The rest of the
quark linear
combinations are in first approximation decoupled.

Further we have to care for approximate CP conservation. Therefore, in order
to approximately diagonalize the strength of coupling to the $g_t$ coupling
blobs, we shall
choose CP-eigenstates. The quarks and the leptons in the usual basis
do not diagonalize CP. We have to consider linear combinations of quarks
and their anti-quarks! This might make the reader worried about superselection
rules according to which one cannot make superpositions of states with different
electric charges for instance. In our approximation of first only including the
$g_t$ Yukawa coupling we have totally ignored the gauge fields,
especially the electromagnetic fields, and thus the gauge symmetry is not
so relevant
in this scheme. The point is rather that, with the pure $g_t$
coupling, we have a theory with CP-invariance in first approximation. Thus,
unless CP should get
spontaneously broken, our first approximation model is totally CP-symmetric and
mass - or here better interaction strength - eigenstates will
be CP-eigenstates, unless they are degenerate. To be concrete, a priori there is
mixing between a quark and the corresponding antiquark
much like that between $K^0$ and its antiparticle $\bar{K}^0$, which is
well-known. In the approximation of ignoring the CP-violation,
it is well known that the eigenstates for the $K^0$-set are $K_1$ and $K_2$,
which are CP-eigenstates. Of course one usually does not care so much for the
corresponding eigenstates of CP among say the $d$ and the $\bar{d}$ quarks. This is
because under normal conditions electric fields can too easily
cause us to measure the charge and thus do not leave the $d$ and $\bar{d}$
system in peace to reveal its true eigenstates with CP-eigenvalues.
However, in the very short time their blobs interact only via the $g_t$
in our effective interactions, electromagnetism can be neglected
and the true eigenstates can thus be of relevance.

Thus, in our approximation scheme, an important ``eigen''-state for say
the strength of the interaction are the CP-eigenstates in the subspace of
top-right and its  CP-transformed left-helicity anti-top,
which we still say has right chirality.
The same sort of CP-eigenstates can be formed from the
antiparticle particle
linear combination based on the left chirality top and similarly on the
left chirality dsb-combination.

In order to make this a bit clearer, we list the eigenstates of this interaction
strength in our
approximation

\begin{eqnarray}
\hbox{\bf The t-like quarks:}&&\nonumber\\
 \hbox{``CP-odd right chirality top''} &=& \frac{1}{\sqrt{2}}(t_R -CPt_R)=\frac{1}{\sqrt{2}}(t_R
-\bar{t}_{\hbox{left helicity}}) \nonumber\\
 \hbox{``CP-even  right chirality top''} &=& \frac{1}{\sqrt{2}}(t_R +CPt_R)=\frac{1}{\sqrt{2}}(t_R
+\bar{t}_{\hbox{left helicity}})\nonumber\\
\hbox{``CP-odd left chirality top''} &=& \frac{1}{\sqrt{2}}(t_L -CPt_L)=\frac{1}{\sqrt{2}}(t_L
-\bar{t}_{\hbox{right helicity}})\nonumber\\
\hbox{``CP-even left chirality top''} &=& \frac{1}{\sqrt{2}}(t_L +CPt_L)=\frac{1}{\sqrt{2}}(t_L
+\bar{t}_{\hbox{right helicity}})\nonumber\\
\hbox{``CP-odd dsb in doublet with $t_L$''}&=&\frac{1}{\sqrt{2}}(dsb_L -CP dsb_L)
=\frac{1}{\sqrt{2}}(dsb_L
-\bar{dsb}_{\hbox{right helicity}})\nonumber\\
\hbox{``CP-even dsb in doublet with $t_L$''}&=&\frac{1}{\sqrt{2}}(dsb_L +CP dsb_L)
=\frac{1}{\sqrt{2}}(dsb_L
+\bar{dsb}_{\hbox{right helicity}})\nonumber\\
\hbox{\bf Higgs-like }&&\nonumber\\
\hbox{CP-even neutral Higgs}&=& \hbox{particle of } Re{\phi_1}\nonumber\\
\hbox{CP-odd neutral Higgs}&=& \hbox{particle of } Im{\phi_1}\nonumber\\
\hbox{CP-even charged  Higgs}&=& \hbox{particle of } Re{\phi_2}\nonumber\\
\hbox{CP-odd charged Higgs}&=& \hbox{particle of } Im{\phi_2}\nonumber
\end{eqnarray}

In addition there are quark combinations, orthogonal to the here mentioned ones,
which completely decouple from  the $g_t$-material
in the first approximation. Firstly all other 5 flavours of right chirality quarks
(other than the top) are of this decoupled type, whether even or odd under CP. Secondly we can form
linear combinations of the dsb quarks orthogonal to the combination
in the SU(2)-doublet with the top and these left chirality combinations
will also decouple.
These particles will of course interact via the Higgses but with
Yukawa couplings which are not competitive with $g_t$.
Of course, in this very first approximation, the leptons are also decoupled.

So, in this approximation, particles which we here called decoupled will have infinitely
long penetration depths. However the penetration depths of the
even CP combinations mentioned above will be very short, in as far as
they can go into the main diagram as replacements for one of the
particles already participating in that diagram. In a medium
with even CP-eigenvalue, the CP-odd combinations
should couple so that the negative interference between the
coupling of their two components would exclude their interaction and
thus give them
infinite penetration depths modulo CP-violation. However, the diagrams
only involving the $g_t$ Yukawa coupling are by themselves CP-conserving
and thus, in the very first approximation, the CP-odd combinations decouple
on the CP-even medium background. In an odd CP-eigenvalue medium, of course,
it is opposite and
the even CP particle has an infinite penetration depth, while the odd
CP particle is absorbed very quickly.

Especially for the purpose of getting an understanding
of how penetration goes in the case of CP-violating amplitudes, such as
for calculating $\epsilon'$, we shall choose to describe
what goes on by imagining that
a quark superposition with its antiquark
can only get absorbed. Then we shall consider scattering as being
replaced by the absorbing medium somehow (later) emitting another
particle thereby simulating a scattering.

Let us in fact make a description of the scattering
due to our effective field theory
terms based on the following picture:

\begin{itemize}
 \item Each external particle for our effective term interaction
``sees'' a blob/medium which is actually composed from the
surrounding media of the other three external particles.
\item This medium carries in a way extended over the whole
medium the information as to whether
it has a positive or a negative CP eigenvalue.
\item We only consider the type of interaction that lets the
particle focused on be absorbed by the medium of the other three.
This means that, once the focused on particle has interacted/is absorbed by
the medium it
disappears and is no longer there. It can only be absorbed once.
\item If the medium has the same CP-eigenvalue as the particle considered,
then the absorption can be very strong and the particle will not reach far.
If, however, the medium and the particle have opposite CP-eigenvalues, then
the absorption as a quark and as an antiquark will interfere destructively
and thus the absorption rate will be only proportional
to CP-violation and thus be very small. Consequently, in the case of opposite
CP-eigenvalues for the particle and the medium, the penetration depth can be large.
\end{itemize}

Among the cases we are interested in, it will be the scatterings or effective
field interactions involving the $\mu$ and $\tau$ leptons that may
penetrate through,
while the left chirality quarks, especially the top quark itself,
would seem liable to be
absorbed already at the surface of the blob/medium.
However, even these left-handed quarks can penetrate deeply
if they have the opposite CP-eigenvalue to that for the medium they pass through.

\subsection{Application of Penetration}

We now want to argue that
a particle that is
in an  eigenstate of
very strong absorbtion rate would only reach the surface of the blob or medium
ball and thus have an appreciably less effective coupling in its
effective interaction.
Without such an effect
our model for the coefficients
of the various effective field theory terms
would give
a far too strong interaction term relevant for the mass splitting by mixing of
the $K_1$ and
$K_2$ in the $K^0$-$\bar{K^0}$ mixing. Our model for the coefficient to a term
with 4, say, just $d$ and $s$ quarks without any CP-breaking would
namely provide
so large a coefficient that even alone it would give a much larger
mass splitting
than found experimentally. In other words our model would be severely falsified
by e.g. $K^0$ $\bar{K}^0$ mixing or by other meson mixings, if we do not modify
the prediction for our anomaly in these mixing cases.

The idea now is that, for just these
unwanted predictions,
we can claim that the particles do not penetrate sufficiently deep for
the rule(s) of our original model to
become true. Indeed the quarks relevant  for these mixings
are supposed to interact by being interpreted as components
of the very strongly interacting dsb-quark antiquark combination corresponding to the
$SU(2)_{weak}$-partner of the top. It gets very quickly absorbed and
we can very reasonably speculate that it only reaches the surface of the
medium or blob discussed above.

So far the data would be well fitted by simply leaving out the
contribution to mixing, in the case
that none of the particles can penetrate because of such strong absorbtion.
One has namely
not seen any anomaly so far in the (CP-invariant) mixings.

However these are the strongest terms according to our original rule
and it does not seem reasonable to assume they should be totally removed.
Rather we would say:

Once a particle hits the blob or medium it will either get absorbed sooner
or later
or it will escape on the opposite side. If it has a very strong absorption rate
it will get absorbed almost immediately on the surface, while if it has a lower
absorption rate it will penetrate deeper and for very low absorption rate
easily
escape out on the opposite side. But even the high rate of absorption particle
gets absorbed and thus must have a higher or equally high chance of
producing a process as the less strongly absorbed particles (in a given state
of the medium w.r.t. CP eigenvalue). So there is a limit as to how strongly we
can modify our model to reduce the effective field theory terms, which have all
particles with strong absorption rate. We, so to speak, cannot make the
coefficients
smaller than an analogous process with a smaller absorption rate.

We will now discuss the important example of how much
the "staying at the surface" effect can bring down our predictions for various
specific processes:

  Let us compare a CP-conserving effective field theory term
with only $d$ and $s$ and $b$ quark or anti quark combinations/superpositions
with a corresponding CP-violating process. Concretely you can think
of the CP-conserving process as giving the term leading to an anomalous
contribution to the $K_1$ to $K_2$ mass difference, while the CP-violating term
could be the term we use to explain the anomaly giving the
anomalous $\epsilon'$:

In the CP-conserving case the quark antiquark superpositions meet a medium
with the same CP-eigenvalue as themselves and thus they get absorbed
extremely fast.
In the CP-violating case the quark anti quark superposition meets a medium
with the opposite CP-eigenstate as its own. Thus in the violating case
the particle penetrates much further into the medium.

But now as a particle that has hit the medium region can only get absorbed
once or escape on the other side, the chance of getting absorbed at all must be
at least as big for the high rate of absorption particle as for the
low rate of absorption one. In this example this means that the
chance for the activity of the CP-conserving process must be at least as big as
that for the CP-violating one.

Let us point out that by accident it could quite likely happen that say the
CP-violating
quark anti quark penetrating through the medium would be absorbed with
high probability
before reaching the other side. In such a case the total
strength of the CP-violating and the CP-conserving processes would be equal
to each other.

If this happens, we would in our now modified model obtain the  prediction,
that the anomalous part with and without CP-violation would be equally strong.

\subsection{Several CP-operators}
\label{severalCP}
We have tended to talk loosely about the CP-operator as if there were a
well-defined
CP-operator at least in the rest frame of a particle. But the truth is that
one can modify the CP-operator by multiplying it by some exponential of a
linear combination of
flavour charge operators. If for instance $CP$ is {\bf a} CP-operator and
the operators $N_f$ for various flavours $f$ are the number operators for the
number
of flavour $f$ quarks minus anti quarks, then we can define an infinity
of other ``CP-operators'' by
\begin{eqnarray}
 (CP)_{new} &=& CP\exp( i \delta_f N_f).
\end{eqnarray}
  Whenever you have a  contribution to the Hamiltonian or a diagram with
only {\bf two}  families, it is possible to choose the freedom
in defining a $CP$ operator, i.e. the phases $\delta_f$ so that
just this Hamiltonian contribution or diagram becomes CP-invariant
under the $CP_{new}$.

If we think of the theory in our above used very crude approximation,
in which we ignore masses in propagators because we ignore the
vacuum Higgs field expectation value and rather consider the dsb-superposition
connected by weak $SU(2)$ to the top as one particle, we have only
involved {\bf one} family, namely the third family. So for this case
it is possible to choose a CP-operator that will be conserved in that
approximation. For such a choice and such an approximation, the
CP-odd superposition would not be absorbed at all in passing through a
CP-even medium.

If we want to consider that there can be CP-violating effects and that
such a CP-odd superposition will not penetrate infinitely deep into
a CP-even medium, then we can either:
\begin{itemize}
 \item Choose a CP-definition that makes some other diagram
than our anomalous one CP-invariant. In the case of our estimate
above of $\epsilon'$ we have interference of our
anomaly with a tree-diagram weak interaction term, involving
of course the two lowest mass families when we think of
the $K^0$ to $\pi\pi$ transition. With such a choice our
anomaly diagram will no longer be CP-invariant.
\item Alternatively we could start decorating our diagrams,
which at first
only have $g_t$ vertices diagrams, by allowing - in a
more accurate approximation - also other smaller Yukawa couplings,
or even transverse $W$'s and $Z^0$'s. With sufficiently
many of these smaller couplings CP-violation would at
the end be unavoidable, in the sense that whatever way we would
choose the definition of the CP-operator at the end we could not
achieve its conservation any more.
\end{itemize}
Using one of these formulations - either a foreign
CP-choice or the inclusion of a next level of corrections -
we can peacefully think of CP as slightly broken, and thus
the CP-odd superposition going through the CP-even
medium would only get a finite penetration depth.

As an exercise we may estimate the strength of absorption of an
odd-CP superposition going through an even CP medium in the two cases:

\begin{itemize}
 \item The Wolfenstein representation \cite{Wolfenstein} of the
Cabibbo Kobayashi Maskawa matrix,
with as real as achievable mixing elements $V$'s,  has only complex items
in the third to first or opposite matrix elements. This corresponds to the fact
that we can choose a CP-operator, which  only gets broken  when these
third to first or opposite matrix elements are involved. So this
Wolfenstein representation gives us a CP-choice of the type that
keeps the tree-diagram dominating the $K^0 \rightarrow \pi\pi$ process
CP-conserving. If we think of using this CP-operator choice
with our non-perturbative $g_t$ alone type of diagram, the coupling dsb-superposition
(in doublet with top) will no longer behave in a CP-conserving way. This
means that a linear combination of this superposition with its
antiparticle superposition will not go totally unabsorbed through
the even CP medium. The Wolfenstein mixing matrix between t and d
has an order of unity phase but is numerically very small of the
order of 1\%. Thus we can think of the odd-CP superposition of the
 dsb-quarks and anti-dsb quarks (in doublet with the top) as
most likely being either ($b$ or $\bar{b}$) or ($s$ or $\bar{s}$)
while there is only a probability of the order of one part in 10000
for it being a $d$ or $\bar{d}$. Only in the latter unlikely case
can it violate the CP as suggested by the Wolfenstein formalism.

This thus suggests that, in this consideration, the odd CP eigenstate
dsb combination interacting will penetrate about 10000 times
further into the CP-even medium than the corresponding even
superposition. If you count that the $V_{td}$ has about equal
real and imaginary parts, we might say that the probability for absorption is even about
a factor of 2 smaller. Thus the penetration should be
about 20000 times deeper than that for the even CP partner.

Without violating the principle that the process amplitude for the
even CP superposition cannot be lower than that for the odd one,
we could at most postulate that the effect of the even CP superposition only
reaching the surface could reduce its amplitude by a factor 20000.

\item Alternatively we could think of choosing at first a CP-definition
leaving our first approximation ``only $g_t$ diagrams'' exactly
CP-invariant, but then introduce smaller corrections by allowing
also diagrams with a few smaller Yukawa couplings involving other
families than the third one. Let us for simplicity choose to keep the
approximation of no vacuum Higgs field and thus masslessness of all the
involved quarks. It is indeed possible if we have in the perturbative
corrections to our main diagrams in addition involved enough
CKM-matrix elements to form a Jarlskog triangle, meaning a combination
of the type
\begin{eqnarray}
 J &=& Im( V_{us}V_{cb}V_{ub}^* V_{cs}^*)
\end{eqnarray}
or any of the nine combinations of analogous
flavours. By considering that these imaginary parts of products
of CKM-matrix elements actually give the areas of the  Jarlskog unitarity triangles,
they have the same values.  Thus, inserting into a diagram
of ours, a set of non-maximal Yukawa couplings so as to deliver
an expression leading to the $J$ makes the diagram develop an imaginary
part, which cannot be removed by changing the
phases built into the proposed CP-operator with which a quark goes into the
corresponding anti-quark - i.e. changing the $\delta_f$'s above.
That is to say such a decorated diagram will develop
an imaginary part, meaning a breaking of CP, for whatever definition of
CP you might choose, being just $J$, the Jarlskog invariant.
Most easily - with biggest value - we shall obtain the appearance
of this kind of expression in the interference terms in the squared amplitude for the
absorbtion of a CP-odd particle superposition passing the even CP medium say.
Thus it will be absorbed by a rate proportional to this very Jarlskog invariant
$J$. In the Wolfenstein representation its value is given by
\begin{eqnarray}
 J &\approx & A^2\lambda^6 \eta,
\end{eqnarray}
where $A=0.82$ and $\eta= 0.34$ are of order unity while
$\lambda =.22$ is the Cabibbo angle.
Thus $J\approx 3.1*10^{-5}$. This is essentially the same as
the factor $1/20000$ estimated above.
So, in both cases, we conclude that
the "only reaching the surface effect" could not
suppress the CP-even on CP-even medium absorption by more than
this factor of $3.1*10^{-5} = \frac{1}{30000}$.
\end{itemize}
It seems we have learned the rule that, whatever the precise definition of the
CP-operator considered, it will turn out that the absorption rate in the
CP-violating case - when medium and particle superposition have opposite CP values -
will be slower than the CP-conserving rate by a factor being just equal to
the Jarlskog invariant $J=3.1*10^{-5}$.
So, measured in units of the penetration depth of the CP-even particle
into a CP-even
medium, the penetration depth of the CP-odd particle into the CP-even medium
becomes the inverse of the Jarlskog invariant $J^{-1}$, thinking at first of
an infinitely deep medium.

\subsection{Only-at-surface suppression.}

The reason for our interest in this penetration of especially the
CP-eigenstates
of quark type is that we want to estimate how the lack of penetration can
modify the effective field theory coupling for a process - say a
dimension = 6 term -
simulating our non-perturbative effects due to a strong coupling $g_t$.
Instead of truly estimating the thickness of the medium we talk about
and thereby the effect, we shall allow ourselves to introduce a second
parameter
to fit in our model. A priori we shall think of this second parameter to be
related
to the depth of the medium/blob relative to the depth into which the
CP-nonviolating
superposition - the even CP particles towards the even CP medium - penetrates
with its very
strong absorption. Let us denote the depth or diameter $D$ of the medium
measured in units
of the penetration depth of the even CP particle on an even CP medium. Then our
modification
of our original model consists in multiplying the effective field theory
coefficient - which is what in the original model we calculate as our first
parameter $K$ multiplied by some suppression factors - by an extra factor
$\frac{1}{D'}$. This extra factor takes care of the fact that a particle, which does not penetrate
except into the surface, must have a smaller effective interaction strength
than if it penetrates and can interact with the whole body of the medium.

\begin{itemize}
 \item If the penetration depth $J^{-1}$ of a CP-odd superposition is longer
than the diameter $D$, i.e. if $J^{-1} > D$, then the particle will
often go through the medium and end up not being absorbed. In this case
the probability for the absorption of the CP-odd particle can be small compared
to its probability for hitting the medium at all. A corresponding CP-even
particle
will however, taking its penetration to be smaller than the diameter, always
be absorbed. Thus the chance of absorption for the CP-odd particle will be smaller
than that for the CP-even one. Consequently we also expect the effective coupling for
the process involving the CP-odd particle to be smaller than that involving the
CP-even one instead. This means that the CP-even generated process, which is suppressed
relative to our original model by a factor $D'$, is actually suppressed by a factor $D$.
The region through which the CP-odd particles can interact with
the medium
is namely $D$ times deeper than the region or depth for the CP-even particle.
So indeed in this case of $J^{-1} > D$ :
\begin{eqnarray}
 D' &=& D.
\end{eqnarray}
\item If on the other hand $J^{-1} < D$ the CP-odd particle gets absorbed
inside the medium before reaching the other side. In this case the ratio of the
penetration depths for the CP-odd to the CP-even particle (still in a CP-even
background medium)
is just the Jarlskog invariant inverted $J^{-1}$. That is to say we now have
the extra
suppression factor
\begin{eqnarray}
 D' &=& J^{-1}.
\end{eqnarray}
\end{itemize}

Thinking of the difficult to estimate (second) parameter $D$ as a random
variable, which we can only guess crudely if at all, we are actually interested in the
distribution of the related parameter $D'$, which is the only one going into
our predictions. One would of course say that in such a
statistical expectation there should be a finite non-zero probability for $D> J^{-1}$, so that even
the odd-CP particle
does not penetrate through the medium but gets absorbed on the way through.
In this
case we saw that $D'= J^{-1}$. So this special value for the suppression
factor $D' = J^{-1}$
has a finite non-zero probability for occurring. So the probability
distribution has
a delta-function peak at this $J^{-1}$ value. If this is not the value for
$D'$ then we must have
\begin{equation}
 1\le D' \le J^{-1}.
\end{equation}

It happens that
our most simple prediction for the
anomaly for the effective field theory term relevant for the $K^0$ $\bar{K^0}$
mixing mass difference was just $3 *10^4$ times larger than the whole
Standard Model prediction for this term, a ratio accidentally(?) equal
to the inverse of the Jarlskog invariant $J^{-1}= 3*10^4$.
Now, according to the above considerations, an especially likely value
for the parameter $D'$ which should correct a process or an effective
field theory Lagrangian term is $D' = J^{-1}$. Taking this especially
likely value $D' =J^{-1}$, the simplest version of our model
for the mixing term relevant for the mass difference would be cut down
by a factor of just $D'=J^{-1} = 3*10^4$. This would make our
corrected anomaly prediction be just of the same order as the Standard
Model prediction in this case. Above in equations (\ref{Bai}, \ref{Experiment})
we saw that the theoretical prediction from the Standard Model for the mass difference was
$5.5 *10^{-12}$ MeV
while experiment gave $3.5 *10^{-12}$  MeV, i.e. theory was
$\sim 50 \%$ too big.

So the deviation between theory and experiment in this mass difference
case is really of the same order of magnitude as the theory prediction itself.
So the best result for our model in the modified form would indeed be that there
were an anomaly in this mass difference and we could, in this case, say that
we predicted its order of magnitude correctly. However, it is of course
o.k. just to say that our prediction for an anomaly in the modified version of
our model just barely manages to be as small as the uncertainty. So our modified
model has no problem with this process of mixing, although the
simple model prediction looked like falsifying our model.


\section{A problem with $\epsilon$}
\label{epsilon}
At first it seems that, even after the inclusion of the
story about the stopping at the surface of the too strongly interacting
eigenstates of particles, our model still has the problem of predicting
an anomaly in $\epsilon$, the parameter of the first CP-violating
experiment by Christensen, Cronin et al.

The problem is that, even including our suppression from particles
that may not penetrate into the bulk of the interacting material,
there is still an anomalous effect of
our non-perturbative type present in the $K^0$ mixing system, making the
transition between $K^0$ and $\bar{K}^0$. Before the
suppression by the penetration effect of the CP-conserving part,
this CP-conserving part would be dominant. However after
this penetration effect suppression, we expect that the remaining anomaly could
have a CP-violation effect of the same order as the CP-conserving one.
Now we already found that the (CP-conserving) shift in the
mass-difference due to our anomaly in fact turns out to be barely visible.
This means that we actually predict that the present deviation of
$\Delta M$ from the Standard Model prediction {\em is} already
an anomaly of our type. But if so, then our about equally large
CP-violating anomaly contributing to $\epsilon$ would be dramatic, appreciably
larger than the Standard Model CP-violation terms!

A priori it therefore looks that such a dramatic anomaly in the first discovered
CP-violation parameter $\epsilon$ would kill our model.

Here, however, we would like to point out that it is in reality not going to be
quite as dramatic:

In fact we shall refer to a relation, derived under
rather general (see e.g. \cite{Nikhef}) conditions, concerning the possible
values for the CP-violation
parameter $\epsilon$:
\begin{eqnarray}
\frac{|\Gamma_{12}|}{|M_{12}|} *\sin(\phi) &=& 2(1 -\frac{|q|}{|p|})\\
&\approx&  4Re(\epsilon). \label{rel}
\end{eqnarray}
Here $|\Gamma_{12}|$ and $|M_{12}|$ are the numerical values
of respectively the off-diagonal matrix element in the mass matrix for
the $K^0$ to $\bar{K}^0$ system from the decay rate and from  the mass.
The angle $\phi$ is related to the argument of the ratio of the two
types of off diagonal
matrix elements $\Gamma_{12}$ and $M_{12}$,
\begin{eqnarray}
\phi &=& arg(-\frac{M_{12}}{\Gamma_{12}}).
\end{eqnarray}
 Calculating e.g. the important box-diagram amplitude from the Standard Model for the indirect
CP-violation as well as for the mass difference, we see that
the $\Gamma$ (and thus especially the off-diagonal $\Gamma_{12}$)
represents the absorptive part, while the dispersive part is
in $M$ having the off-diagonal $M_{12}$.

We can only obtain an absorptive part via the $K^0$ decay open
channels and thus it only comes from diagrams involving such a
low energy scale that energies of the order of the K-meson mass
are achievable. The absorptive part from our anomalous term, for which the supposed mass scale is
very high of the order of a half TeV, will be totally
negligible. Our anomaly will in practice only give a {\bf dispersive}
part and thus can only contribute to $M_{12}$ and completely
negligibly to $\Gamma_{12}$.

In this light we see, that provided we have the relation
(\ref{rel}) even when the anomaly is switched on,
it is {\em only} $M_{12}$ but not $\Gamma_{12}$ that can be
changed by the anomaly term. Thus
the real part $Re(\epsilon)$ of the $\epsilon$-parameter
cannot get dramatically larger than the Standard Model value,
since $\sin \phi$ can at most go up to unity, and the
numerator $|\Gamma_{12}|$ cannot be changed by even a huge
anomaly effect. Order of magnitudewise the anomaly prediction
for $Re(\epsilon)$ is bounded from above by the Standard Model value.

Can we even get an effect of the anomaly for $\epsilon$ downwards
because of the anomaly effect in $|M_{12}|$?

It cannot be very dramatic, because the numerical value of the off-diagonal matrix element
$|M_{12}|$ already contains the CP-conserving
part, which gives rise to the mass difference $\Delta M$
between $K_L$ and $K_S$. So the relative anomaly content in
$|M_{12}|$ should only be of the same order as the relative anomaly content
in the mass difference. Although this mass difference could well
be anomalous by a factor 2 say, there is no possibility for much more.
But remember such a factor 2 was indeed what we a priori predicted.

So indeed our prediction for an anomaly in $\epsilon$ is after all
rather stabilized both against going up and going down.
It is thus really not a killing of our model that the
Standard Model calculation for the $\epsilon$ - in our
point of view happens to - agree(s) with experiment.

We only have to suggest that the overall value of the
anomaly fell a tiny bit to the low side in the $K^0$ $\bar{K}^0$ mixing case,
so that even the anomaly in the mass difference - which we may
pretend to see - is a bit low. Thus the change in $\epsilon$ by the anomaly
gets to be of some order like say 30\% and so is not
really observable at present.

\subsection{Physics of our keeping anomaly in $\epsilon$ down}

It is of course very crucial for rescuing our non-perturbative model
for the anomalies that we understand how we avoided obtaining
a huge anomaly for the CP-breaking parameter $\epsilon$.

The physics behind this may be understood by thinking about the
extreme case where one had a purely dispersive CP-breaking effect,
coming from diagrams that could be completely represented
by a dimension 6 effective field operator  - like our model
for the anomaly. The effective Hamiltonian for such a term
would at first not have to have a real coefficient, in as far as
it is a term adding say two strangeness units and then from hermiticity
there should be a corresponding term removing two strangeness units.
These two corresponding terms would of course have complex conjugate
coefficients. But each of them can have a complex coefficient.

However now there is the freedom, discussed above in subsection \ref{severalCP},
that one can construct several $CP$ operators, by supplementing
one with exponentiated $i$ times flavour charge operators.
When we consider a transition matrix element between a $K^0$ and a
$\bar{K}^0$ state, such as $M_{12}$, then of course its phase can be
changed by modifying the phase convention for the states with
strangeness. Such a phase change is achieved for
a CP-operator, if it is written in terms of $K_1$ and $K_2$, when this
CP-operator is changed by being multiplied by a factor $\exp(iaS)$, where
$S$ is the strangeness operator:

In fact, if we define the phases on the  $|K^0>$ and $|\bar{K}^0>$ by
defining them from
\begin{eqnarray}
|K^0> &=& \frac{1}{\sqrt{2}} ( |K_1> + |K_2>)\\
   |\bar{K}^0> &=& \frac{1}{\sqrt{2}} ( |K_1> - |K_2>)\\
\hbox{ and}&&\nonumber \\
``CP''|K_1>  &=& |K_1>\\
``CP'' |K_2> &=& -|K_2>\\
\hbox{we could take}&&\nonumber \\
``CP'' &=& CP\\
\hbox{or we could take} &&\nonumber\\
``CP'' &=& CP \exp(iaS).
\end{eqnarray}
In this way we can shift the relative phase of these states  $|K^0>$ and $|\bar{K}^0>$
by adjusting the choice of the parameter $a$.

So we can now take such a definition of the
CP-operator $``CP''$  that the transition matrix element
$M_{12}$ becomes {\em real}. In this way we can make the
purely dispersive matrix for the development of the  $|K^0>$ plus
$|\bar{K}^0>$
system $``CP''$-invariant. So in such a purely dispersive case
- i.e. with no decay - there will be no violation of
the chosen ``CP'' and thus there should be no
decay of $K_L$, which will actually be $K_2$, into 2 pions.
Thus indeed in such a case $\epsilon =0$.

This means that one can only achieve a non-zero
$\epsilon$ gradually as one screws up  the
absorptive part. But, as long as the absorptive part is in some sense
small, $\epsilon$ will be kept small in the same
sense, even if there are huge dispersive terms.

Now if, as we suggest in our model, the anomaly only influences the
dispersive part and that only by a relative magnitude
close to being of order unity, but possibly a little less
say of the order of 30\%, then the anomaly in $\epsilon$ will
also only be of this order. This is true even if it corresponds to a
shift in the $M_{12}$ by about 30\% and this anomalous shift
has a quite different phase from the term it adds to, so that one
would at first think it gave a huge CP-violation.

\section{Conclusion}
\label{Conclusion}
We have looked at 5 ``anomalies'' meaning (small) deviations from
the predictions of the Standard Model, and proposed that they are indeed
due to {\bf non-perturbative effects} rather than to genuine ``new
physics''. We have fitted them order of magnitudewise by a somewhat
arbitrary model based on the thought that, because of the
rather large size of the top-Yukawa-coupling $g_t$, there are important Feynman
diagrams of extremely/infinitely high order. We basically think of the
extremely high order diagrams as just being modified by arranging them
to have a few external lines. For practical purposes, in this philosophy,
we then construct some effective field theory terms
actually of dimension 6, which simulate the effect of the
infinitely high order diagrams. Because we cannot evaluate
the sum of the infinite order diagrams, we are forced to introduce
one over-all scale parameter, which we call $K$.

Remarkably enough we obtain by the fitting of $K$ to the five
different - and indeed very different in nature - anomalies
order of magnitudewise the same value for $K$ ! This should
be considered a great success of our idea, and is at least somewhat
evidence for these anomalies indeed being due to non-perturbative effects.

Even more surprisingly the different fitted values only deviate from each other by
up to 20 \%, except for the anomaly in the muon magnetic moment.
However, even for that case, we may by somewhat untrustable treatments of the factors of 2
obtain a value that could also agree within the 20\%.
Strictly speaking there is possibly no reason that we should expect
our model only to work order of magnitudewise. So we might hope in the
future to convince ourselves and others that there is a way to make the
treatment
of the model such that one should expect a 20 \% accuracy.
After all our procedure is only to {\em modify} Feynman
diagrams, that a priori are (almost) the same for all the 5
considered anomalies.

\subsection{PREdictions}

Our model at first seemed to have a very severe problem:

It predicted far too large anomalies in $K^0\bar{K}^0$ mixing and other
similar meson anti-meson mixings. With the purpose of avoiding such
predictions disagreeing
with experiment, we invented a physical mechanism
of {\bf the penetration of too strongly coupling particles getting stopped}, when
in an interaction they are about to penetrate into each other (into the cloud
 of top quarks and Higgses around the other
particle). But even this invented story only barely solves the problem,
so we predict that we are very close to finding some anomalies in the mixing
of meson systems. In fact the mass difference between $K_L$ and $K_S$
deviates from the Standard Model prediction by about a couple
of standard deviations, and that would fit well with our estimate
of an anomaly in this mass difference. We, so to speak,
predict this should turn out to be a true anomaly.

For the $CP$-breaking $\epsilon$ we at first seemed to get an
intolerably large anomaly, completely dominating the Standard Model
contribution. However, because our anomaly is completely
dispersive and cannot change the absorptive part - $\Gamma_{12}$
the transition width - the anomaly in $\epsilon$ cannot be so terribly
large after all. It is, however, still a tiny bit of an accident
that one has not seen any anomaly in $\epsilon$ yet, but now it is not
an outrageous accident.

A rather clean prediction from our model is that there shall be an
even dominating anomaly in the decay channel analogous to the
$R(K)$ and $R(K^*)$ case discussed but with $\tau$ replacing the
$\mu$, i.e. in the channel:
\begin{eqnarray}
B\rightarrow K^{(*)}\tau^+\tau^-.
\end{eqnarray}
We namely get stronger and stronger anomalies the bigger the Higgs Yukawa
coupling for the lepton.

\subsection{The inspiration, old ideas on bound state etc.}

The idea of non-perturbative effects due to the largeness of the
top-Yukawa coupling $g_t$ can be considered as going back to
our earlier work, where we speculated about the existence of
bound states of 6 top + 6 anti top quarks \cite{coincidence}, caused by Higgs as well as
gluon exchange. Such bound states are precisely the type of
non-perturbative effects we could have in mind. But in the present
work we took such an abstract view on the non-perturbative effects that
they could have been many other configurations - new vacuum etc. -
or just high order diagrams being important without any extra effects.

We estimated that the theoretical border for non-perturbative
effects, taking into account the number of different components
of Higgs-bosons and top-quarks interacting via the top-Yukawa coupling, is
indeed very close to the experimental value of the top Yukawa coupling
$g_t = 0.935$. So indeed theoretically the coupling is on the borderline to be
``strong''.

Also our long discussed ``Multiple point principle'' \cite{MPP}
may have a place by explaining why the Yukawa coupling is
just on the borderline of being strong (if indeed it is).

 \subsection{Outlook}

Encouraged by the only 20 \% deviations between the
different $K$-values fitted, it would be natural to attempt
to redo our calculations fitting the $K$-parameter in such a way
that such an accuracy would make sense.

A slightly related
 possible future project would be
to estimate the {\bf sign} of the effects. Indeed the signs vary
from case to case, from anomaly to anomaly, in a way that
would be hard even to give any meaning to, in the light of the
different anomalies being so very different.
But we think the first progress in estimating the sign could be to argue that
the interaction between the muon and the hadronic matter in the
shrunk proton puzzle has the sign corresponding to there being an
attraction in the anomaly-term. This is very natural for
a high mass scale effect (of a scalar shape, meaning no
spin coming in in the supposed bound state).

\section{Acknowledgement}
It should be mentioned that we have made appreciable progress on this
work since the workshop in Corfu last year, where we basically only
had 3 out of the now 5 anomalies under consideration.

One of us H.B. Nielsen wishes to thank the Niels Bohr Institute for being allowed
to stay there as emeritus, and also for some economic support, although not
sufficient for it all, as well as thanks to the conference organization,
George Zoupanos et al.

Most of the progress was made while Colin Froggatt visited the Niels Bohr
Institute and he thanks his friends - Chand and Wendy Sapru - for accommodation
during this period. He also wishes to acknowledge hospitality and
support from Glasgow University and the Niels Bohr Institute.

Also we wish to acknowledge an important question that was raised  at the conference in
Tallin in 2018.

\end{document}